\documentclass[english,11pt]{article}
\usepackage{multicol}
\usepackage{amsmath}
\usepackage{amssymb}
\usepackage{bbold}
\usepackage{mathrsfs}
\usepackage{breqn}
\usepackage{graphicx}
\usepackage[table]{xcolor}
\usepackage{accents}
\usepackage{float}
\usepackage{color}
\usepackage[T1]{fontenc}
\usepackage[utf8]{inputenc}

\usepackage[colorlinks=true,urlcolor=blue,linkcolor=blue,citecolor=red]{hyperref}

\usepackage[
backend=bibtex,
style=numeric,
maxnames=3,
minnames=1,
sorting=none,
doi=false,
url=false,
isbn=false
]{biblatex}

\renewbibmacro*{journal}{%
  \iffieldundef{shortjournal}
    {\printfield{journaltitle}}
    {\printfield{shortjournal}}}

\def\sr{{\text{\tiny)}}}
\def\sl{{\text{\tiny(}}}
\def\ar{{\text{\tiny]}}}
\def\al{{\text{\tiny[}}}
\def\n{{\text{\tiny 0}}}
\def\on{{\text{\tiny 1}}}
\def\tw{{\text{\tiny 2}}}
\def\tr{{\text{\tiny 3}}}
\def\h{{\text{\tiny h}}}

\begin{document}
\begin{center}
{\Large \bf On non-vacuum black holes in new general relativity \\ }
\vskip 0.7cm
{\bf D. F. L\'opez\footnote{Corresponding author: diego.lopez@dal.ca \\}, A. A. Coley, B. Yildirim}
\vskip 0.2cm
{\it Department of Mathematics and Statistics\\ 
Dalhousie University\\
Halifax, Canada}\\
\vskip 1cm
\begin{quote}
{\bf Abstract.}~{\small New general relativity (NGR) is a torsion-based modification of general 
relativity whose Lagrangian depends on three free parameters, $(c_{a}, c_{v}, c_{t})$. 
A subset of the parameter space is physically admissible, namely that which 
simultaneously ensures ghost-freedom, propagation of a spin-2 mode, and a consistent 
Newtonian limit. In this work we analyze static and spherically symmetric 
configurations in NGR, both in vacuum and in the presence of a perfect fluid and an 
electromagnetic field, under the assumption of the existence of a local black-hole 
horizon. We find that the mere existence of such configurations forces the free 
parameters into regions associated with known pathological models: theories that 
either contain ghost instabilities, do not propagate a spin-2 mode, or lack a 
Newtonian limit. The remaining geometries are regular at the horizon, so the 
obstruction is not a breakdown of the geometry but a breakdown of the underlying 
theory. We therefore conclude that, within the class of models examined, NGR does not 
admit physically meaningful non-trivial black holes distinct from those of the 
teleparallel equivalent of general relativity.
}
\end{quote}
\end{center}
\vskip 1.0cm

%
%
\section{Introduction}

The primary geometrical object in teleparallel gravity (TG) is the torsion tensor, which is constructed from a coframe and a curvature-free, metric-compatible spin connection. While in general relativity (GR) gravity is a purely geometric effect encoded in the curvature of the Levi-Civita connection, in TG the gravitational interaction is attributed to torsion. The teleparallel equivalent of general relativity (TEGR) is a  particular subclass of TG in which the action is built from the torsion scalar \(T\). This theory is locally dynamically equivalent to GR, sharing the same field equations (FE) and classical  predictions, including the existence of an analogue of the Schwarzschild  solution~\cite{aldrovandi2012teleparallel, krvsvsak2019teleparallel}.

However, in the Schwarzschild-like solution of TEGR, the behavior of geometrical invariants differs markedly from that in GR. In GR, curvature singularities typically occur only at the origin of the radial coordinate. However, torsion scalar invariants in TEGR, and particularly  the scalar \(T\), also diverge at the Schwarzschild horizon. Indeed, in the static, spherically symmetric (SSS) vacuum case, the FE of TEGR yield the Schwarzschild metric together with a tetrad defined up to local Lorentz transformations. In this setting one finds~\cite{obukhov2003metric, coley2025black}
\begin{equation}
T=-\frac{4\left(M-r+\sqrt{r(r-2M)}\right)}{r^{2}\sqrt{r(r-2M)}}.
\end{equation}
As \(r \to 2M\), the torsion scalar \(T\) diverges for generic choices of the free functions in the spin connection.\footnote{It has been recently observed that particular choices of these functions may remove such divergences; a detailed analysis will be presented in a forthcoming work.} It should be emphasized that, despite its equivalence to GR at the level of FE, the TEGR case is still not fully understood from the perspective of its invariant geometrical structure and its interpretation~\cite{coley2025black, lopez2025black}.

A generalization of TEGR is \(F(T)\) gravity, in which \(F\) is an arbitrary twice-differentiable function of the torsion scalar that appears in the action~\cite{ferraro2007modified, krvsvsak2019teleparallel}. When the geometrical framework of \(F(T)\) theory is formulated in a gauge-invariant manner, the resulting FE are fully Lorentz covariant~\cite{krvsvsak2019teleparallel}. In this formulation, there always exists a frame--spin-connection pair for which the spin connection vanishes identically (the so-called ``proper'' frame)~\cite{aldrovandi2012teleparallel, krvsvsak2019teleparallel}. A general family of SSS vacuum solutions in \(F(T)\) gravity has been presented in~\cite{coley2024spherically, van2024teleparallel, bahamonde2021general}. Recently~\cite{coley2025black} it was shown that SSS vacuum spacetimes in \(F(T)\) gravity in which a putative local horizon (LH) exists necessarily exhibit a divergence of the torsion scalar \(T\) at that location. As a consequence, such spacetimes cannot be interpreted as black hole solutions.

New general relativity (NGR) is a torsion-based modification of GR defined by including additional irreducible torsion scalars in the action, parameterized by two free constants (with a third parameter fixed by normalizing the effective gravitational constant). TEGR is recovered as a particular case, corresponding to a specific linear combination of these irreducible torsion invariants~\cite{hayashi1979new, hayashi1981addendum, aldrovandi2012teleparallel}. In its original non-covariant formulation~\cite{hayashi1979new, hayashi1981addendum, hayashi1990static}, NGR admits an exact spherically symmetric vacuum solution that is known to describe a geometry with a singular horizon~\cite{kawai1990singularities, obukhov2003metric, lopez2025black}. In a proper tetrad (i.e., a zero-spin-connection gauge), this behavior admits an intuitive explanation: the proper tetrad consists of covariantly constant vectors, and as one approaches the horizon, the time-like Killing vector tends toward a light-like direction, necessarily inducing a singularity~\cite{golovnev2024degrees}.

Using a fully invariant formulation, it was subsequently shown that general SSS vacuum solutions of NGR yield torsion invariants with two singularities~\cite{kawai1990singularities, obukhov2003metric, golovnev2024static, lopez2025black}. Assuming the existence of a LH, it was demonstrated that all physically viable NGR models inevitably exhibit divergences in torsion scalars at that horizon. This singular behavior prevents these teleparallel geometries from being interpreted as black hole spacetimes~\cite{lopez2025black}. A comprehensive classification of solution branches satisfying both the antisymmetric field equations (AFE) and the symmetric field equations (SFE) was presented in~\cite{lopez2025black}. With the exception of two cases that are essentially equivalent to TEGR~\cite{obukhov2003metric}, all remaining branches are unphysical. The unfavorable features of these models include:
\begin{itemize}
\item[(i)] The presence of ghost instabilities~\cite{golovnev2024gravitational, 
bahamonde2025revisiting}. The one-parameter Hayashi--Shirafuji (H\&S) 
model~\cite{hayashi1979new, hayashi1981addendum, hayashi1990static} has often been favored 
based on claims of ghost freedom at linear level~\cite{jimenez2020non, bahamonde2025revisiting}.

\item[(ii)] The absence of propagating spin-2 degrees of freedom, rendering the theory 
incapable of describing gravitational waves~\cite{golovnev2024gravitational, 
bahamonde2025revisiting}.

\item[(iii)] The absence of a consistent Newtonian and/or post-Newtonian limit, which leads to 
incompatibility with solar-system tests~\cite{hayashi1979new, hayashi1981addendum, 
jimenez2020non}. When \(c_v=0\), the GR limit cannot be recovered and the model is therefore 
unphysical. For \(c_v\neq 0\), one may impose the normalization \(b_{2}=2+3b_{1}\)~\cite{lopez2025black}.
\end{itemize}

Despite these shortcomings, particularly at the quantum level, NGR can still be meaningfully employed in classical phenomenology, especially in cosmological applications. Indeed, the most general NGR theory appears to be viable and may represent the most promising framework, as it possesses a well-defined number of degrees of freedom and exhibits robust physical modes~\cite{golovnev2024gravitational, golovnev2024static, golovnev2024degrees, bahamonde2025revisiting}. In this paper, we investigate whether NGR admits non-trivial, non-vacuum black hole solutions. Focusing on SSS non-vacuum spacetimes, we show that if a geometry with a putative horizon exists, then all NGR models that do not reduce to TEGR inevitably exhibit several of the previously identified unfavorable physical features. The vacuum case has been fully characterized in~\cite{lopez2025black}; the present work extends that analysis to non-vacuum settings.

\subsection{Black holes} \label{BH}

We shall assume throughout that black holes are spacetime geometries admitting a horizon that shields an interior spacetime singularity.\footnote{While this is not a universally accepted definition, it is sufficient for the purposes of our analysis; the results obtained remain valid regardless of the detailed nature of the interior region, including the case of so-called singularity-free black holes.} For example, the surface $r_{s}=2M$ constitutes the horizon of the Schwarzschild manifold. This is a ``global'' event horizon; however, since its definition requires global information about the entire future evolution of the spacetime, it is more useful in dynamical situations to characterize horizons ``locally'' by means of an apparent horizon (AH)~\cite{hawking1973large, ashtekar2002dynamical, ashtekar2003dynamical}.

An AH is a marginally outer trapped surface, defined by the following set of conditions~\cite{ashtekar2002dynamical, ashtekar2003dynamical, lopez2025black}:
\begin{equation}\label{AHD}
\theta_{\sl \ell \sr} = 0 \, , 
\qquad 
\theta_{\sl n \sr} < 0 \, , 
\qquad 
\Delta \theta_{\sl \ell \sr} < 0 \, .
\end{equation}
These conditions specify, respectively, that the expansion of outgoing null rays vanishes ($\theta_{\sl \ell \sr}=0$), that the ingoing null congruence remains converging so that the surface encloses a properly trapped region ($\theta_{\sl n \sr} < 0$), and that the outgoing expansion decreases when moving inward along the ingoing null direction ($\Delta \theta_{\sl \ell \sr} < 0$), which provides a stability criterion. In the spherically symmetric case, an AH coincides with a geometric horizon (GH), which can alternatively be defined in a fully coordinate-independent manner through the vanishing of certain Cartan invariants~\cite{coley2018identification, coley2017geometric, brooks2018cartan}.

Therefore, a necessary condition for the existence of an AH at $r=r_h$ is that the outgoing expansion scalar satisfies $\theta_{\sl \ell \sr}(r_h)=0$~\cite{lopez2025black}. We refer to this local condition as defining the LH. This local, coordinate-independent characterization replaces the phenomenological condition $g_{tt}=0$ typically used in SSS coordinates, and is particularly suited to the study of teleparallel geometries, where the same characterization applies since matter fields couple minimally to the metric, as in GR~\cite{lopez2025black}.


\section{SSS teleparallel geometry}

Using Latin indices (\(a=1,2,3,4\)) to label tangent-space components and Greek indices (\(\mu = t,r,\theta,\phi\)) to denote spacetime coordinates, teleparallel geometry is formulated in terms of a tetrad field \(h^{a}{}_{\mu}\) and a metric-compatible, flat spin connection. The latter can be written as \(\omega^{a}{}_{b\mu} = \Lambda^{a}{}_{c}(x)\,\partial_{\mu}\Lambda_{b}{}^{c}(x)\), so that the geometry exhibits non-vanishing torsion while maintaining identically vanishing curvature~\cite{aldrovandi2012teleparallel, krvsvsak2019teleparallel}. Metric compatibility further implies \(\omega_{(ab)\mu}=0\).

The tetrad satisfies the orthogonality relations \(h^{a}{}_{\mu}h_{a}{}^{\nu}=\delta^{\nu}_{\mu}\) and \(h^{a}{}_{\mu}h_{b}{}^{\mu}=\delta^{a}_{b}\), and relates the spacetime metric \(g_{\mu\nu}\) to the Minkowski metric \(\eta_{ab}=\mathrm{diag}(-1,1,1,1)\) through
\begin{equation}\label{mth}
g_{\mu\nu}=h^{a}{}_{\mu}\, h^{b}{}_{\nu}\,\eta_{ab}.
\end{equation}
Imposing the tetrad postulate determines the teleparallel connection and the associated torsion tensor is given by~\cite{aldrovandi2012teleparallel, krvsvsak2019teleparallel}:
\begin{equation} \label{cone}
\Omega^{\rho}{}_{\nu\mu} 
= h_{a}{}^{\rho} \left(\partial_{\mu} h^{a}{}_{\nu} + \omega^{a}{}_{b\mu} h^{b}{}_{\nu}\right), 
\qquad 
T^{\sigma}{}_{\mu\nu} = 2\,\Omega^{\sigma}{}_{[\nu\mu]} .
\end{equation}

The torsion tensor $T^{\sigma}{}_{\mu\nu}$ admits a decomposition into three Lorentz-irreducible parts: a vector, an axial vector, and a purely tensorial component~\cite{hayashi1979new, aldrovandi2012teleparallel}:
\begin{equation}\label{TorDec}
\mathscr{V}_{\mu} = T^{\nu}{}_{\nu\mu}, 
\qquad
\mathscr{A}_{\mu} = \frac{1}{6}\,\varepsilon_{\mu\nu\rho\sigma} T^{\nu\rho\sigma}, 
\qquad
\mathscr{T}_{\sigma\mu\nu} 
= T_{\sl\sigma\mu\sr\nu} 
+ \frac{1}{3}\left(g_{\sigma\al\nu}\mathscr{V}_{\mu\ar} 
+ g_{\mu\al\nu}\mathscr{V}_{\sigma\ar}\right).
\end{equation}
These components can be combined to construct the torsion scalar,
\begin{equation}\label{TS}
T = \frac{3}{2}\,\mathscr{A} 
    - \frac{2}{3}\,\mathscr{V} 
    + \frac{2}{3}\,\mathscr{T},
\end{equation}
where
\begin{equation}\label{AVTS}
\mathscr{A} = \mathscr{A}^{\mu}\mathscr{A}_{\mu},
\qquad
\mathscr{V} = \mathscr{V}^{\mu}\mathscr{V}_{\mu},
\qquad
\mathscr{T} = \mathscr{T}^{\sigma\mu\nu}\mathscr{T}_{\sigma\mu\nu}.
\end{equation}

These scalars play an important role in the formulation of teleparallel Lagrangians, thereby defining a wide class of teleparallel theories. Assuming minimal coupling of matter to the metric, test particles follow force-like equations of motion that are dynamically equivalent to the geodesic equations of GR~\cite{aldrovandi2012teleparallel}. Consequently, the behavior of null congruences can be analyzed in complete analogy with the metric formulation, using the expansion scalars associated with the outgoing and ingoing null directions~\cite{poisson2004relativist, lopez2025black}. In the teleparallel framework, these expansions take the form
\begin{equation}\label{expansionf}
\theta_{\sl\ell\sr}
= \nabla_{\mu}\ell^{\mu}
+ K^{\sigma\mu}{}_{\mu}\,\ell_{\sigma},
\qquad
\theta_{\sl n\sr}
= \nabla_{\mu}n^{\mu}
+ K^{\sigma\mu}{}_{\mu}\,n_{\sigma},
\end{equation}
where \( \nabla_{\mu} \) denotes the covariant derivative with respect to the teleparallel connection \eqref{cone}, $K_{\sigma\mu\nu} = T_{\al\mu\sigma\ar\nu} + \tfrac{1}{2} T_{\nu\sigma\mu}$ is the contortion tensor, and \( \ell^{\mu} \) and \( n^{\mu} \) denote the outgoing and ingoing null vectors, respectively. These vectors satisfy the normalization and affine geodesic conditions~\cite{poisson2004relativist}
\begin{equation}\label{lncond}
\ell^{\mu}\ell_{\mu}
= n^{\mu}n_{\mu}
= 0, \qquad
\ell^{\mu}n_{\mu}
= -1, \qquad
\ell^{\nu}\nabla_{\mu}\ell_{\nu}
= n^{\nu}\nabla_{\mu}n_{\nu}
= 0.
\end{equation}
The expansion scalars in~\eqref{expansionf} are instrumental in the identification of  marginally outer trapped surfaces~\cite{poisson2004relativist, lopez2025black}. As discussed in Subsection~\ref{BH}, such regions are characterized by the conditions~\eqref{AHD}, where  the first condition, $\theta_{\sl\ell\sr}=0$, provides a necessary criterion for the existence of a LH.

With these geometric elements established, the analysis of SSS configurations requires specifying the tetrad and spin connection compatible with the corresponding symmetry group. Working in coordinates $x^{\mu}=(t,r,\theta,\phi)$, the symmetry generators of the affine frame, namely those of the three-dimensional spherical symmetry group together with the time-translation generator $\partial_{t}$, determine the form of the tetrad~\cite{coley2020symmetry, mcnutt2023frame, coley2024spherically, van2024teleparallel}:
\begin{equation}\label{TetradSS}
\boldsymbol{h} =
\begin{pmatrix}
A_1 & 0 & 0 &0\\
0 & A_2 & 0 & 0 \\
0 & 0 & r & 0 \\
0 & 0 & 0 & r\,\sin \theta
\end{pmatrix} \, ,
\end{equation}
where $A_{1}=A_{1}(r)$, $A_{2}=A_{2}(r)$, and, provided $A_{3}(r)$ is not constant, the coordinate freedom has been used to fix $A_{3}(r)=r$. This symmetry-adapted frame leads to the most general SSS metric-compatible connection, whose non-vanishing components are~\cite{coley2020symmetry, mcnutt2023frame, coley2024spherically, van2024teleparallel}:
\begin{equation}\label{SpinV} 
	\begin{gathered}
\omega_{133}=\omega_{144}= \cos\chi \sinh\psi /r \, , \quad 
\omega_{134}=\omega_{413}=\sin\chi \sinh\psi /r  \, ,\\[1.5ex] 
\omega_{234}=\omega_{423}=\sin\chi \cosh\psi/r \, , \quad
\omega_{233}=\omega_{244} = \cos\chi\cosh\psi/r\,,\\[1.5ex]  
\omega_{212}= \psi'/A_{2} \, , \quad
\omega_{432}= \chi'/A_{2} \, , \quad
\omega_{434}= \cot\theta/r \, ,
\end{gathered}
\end{equation}
where $\chi=\chi(r)$, $\psi=\psi(r)$. With this choice, the geometry is fully determined by the four arbitrary functions $A_{1}$, $A_{2}$, $\chi$, and $\psi$, which together uniquely specify the teleparallel geometry~\cite{coley2020symmetry, mcnutt2023frame, coley2024spherically, van2024teleparallel}. Under an appropriate local Lorentz transformation, the tetrad~\eqref{TetradSS} and spin connection~\eqref{SpinV} can be mapped to the so-called proper frame, in which ${\omega'}^{a}{}_{b\mu}=0$ and ${h'}^{a}{}_{\mu}=\Lambda^{a}{}_{b}h^{b}{}_{\mu}$, yielding a completely equivalent representation of the same geometry~\cite{van2024teleparallel}. 

The scalars introduced in Eq.~\eqref{AVTS} play an important role in the investigation of SSS teleparallel geometries. Since these invariants appear explicitly in teleparallel Lagrangians, their behaviour encodes the geometric properties of the arbitrary functions $A_{1}(r)$, $A_{2}(r)$, $\chi(r)$, and $\psi(r)$. Substituting the tetrad~\eqref{TetradSS} and spin connection~\eqref{SpinV} into Eq.~\eqref{AVTS} yields
\begin{subequations}\label{TScalars}
\begin{dmath}
\mathscr{A} = -\left(\frac{4\sin\chi}{3r}\right)^{2} - \frac{16\cosh\psi\, [\cos\chi]'}{9rA_{2}} - \left(\frac{2\chi'}{3A_{2}}\right)^{2}, 
\end{dmath}
\vskip -0.3cm
\begin{dmath}
\mathscr{V}= \frac{4\cos^{2}\chi}{r^{2}} +\frac{4\cos\chi}{rA_{2}}\left([\ln(r^{2}A_{1})]'\cosh\psi + [\cosh\psi]'\right)+ \left(\frac{[\ln(r^{2}A_{1})]'}{A_{2}}\right)^{2}
- \left(\frac{\psi'}{A_{2}}\right)^{2},
\end{dmath}
\vskip -0.2cm
\begin{dmath}
\mathscr{T} = \frac{1}{r^{2}} + \frac{[\ln(A_{1}/r)]'}{A_{2}}\left(\frac{[\ln(A_{1}/r)]'}{A_{2}} - \frac{2\cos\chi\cosh\psi}{r}\right)- \frac{2[\cos\chi\cosh\psi]'}{rA_{2}}
+ \frac{(\chi')^{2} - (\psi')^{2}}{A_{2}^{2}} .
\end{dmath}
\end{subequations}

For a teleparallel geometry to represent a well-defined black hole, all torsion invariants~\eqref{TScalars} entering the teleparallel Lagrangian must remain finite at the LH.\footnote{The Lagrangian density of teleparallel theories is defined only at points where the torsion invariants remain finite; divergence at the LH would therefore exclude the horizon and its interior from the manifold~\cite{coley2025black}.} In the SSS case under consideration, the location of this LH can be determined from the roots of~\cite{lopez2025black}
\begin{equation}\label{theAH}
\theta_{\sl\ell\sr}=-\theta_{\sl n\sr}=\frac{\sqrt{2}}{rA_{2}},
\end{equation}
where we have employed the null vectors compatible with the conditions 
in~\eqref{lncond}, namely~~\cite{coley2020symmetry}
\begin{equation}\label{null}
\boldsymbol{\ell}
=\frac{1}{\sqrt{2}}\!\left(\frac{1}{A_{1}},\frac{1}{A_{2}},0,0\right),
\qquad
\boldsymbol{n}
=\frac{1}{\sqrt{2}}\!\left(\frac{1}{A_{1}},-\frac{1}{A_{2}},0,0\right).
\end{equation}
This particular choice of null vectors is symmetric under both time reversal and radial reflection. As a consequence, the expansion scalars in~\eqref{theAH} have equal magnitude and opposite sign, and therefore their roots coincide.

\section{NGR}

The class of theories known as NGR constitutes the first and simplest modification of TEGR explored in the literature~\cite{hayashi1979new, hayashi1981addendum}. In NGR, the coefficients associated with the scalars obtained from the irreducible parts of torsion are treated as three independent free parameters, \( (c_{a}, c_{v}, c_{t}) \), which are subject to experimental constraints from solar-system tests, gravitational-wave observations, and PPN analyses~\cite{lopez2025black, bahamonde2025revisiting}. These parameters allow for deviations from the TEGR Lagrangian, which depends only on the torsion scalar \(T\) and corresponds to the specific values \( (c_{a} = 3/2,\, c_{v} = -2/3,\, c_{t} = 2/3) \) appearing in~\eqref{TS}. Allowing these coefficients to vary is expected to enable NGR to incorporate corrections to the standard gravitational predictions provided by TEGR.

The most general NGR Lagrangian density is given by~\cite{hayashi1979new, bahamonde2025revisiting}
\begin{equation}\label{Lngr}
\mathcal{L} = c_{a}\,\mathscr{A} + c_{v}\,\mathscr{V} + c_{t}\,\mathscr{T}  \, ,
\end{equation}
where $\mathscr{A}$, $\mathscr{V}$, and $\mathscr{T}$ denote the three independent quadratic torsion invariants, as defined in~\eqref{AVTS}.

The corresponding NGR action, including a matter Lagrangian density $\mathcal{L}_{m}$, is defined as
\begin{equation}\label{S}
\mathcal{S}
= \int h \left( \kappa\,\mathcal{L} + \mathcal{L}_{m} \right) d^{4}x \, ,
\end{equation}
where $h$ denotes the determinant of the tetrad field and $\kappa$ is the gravitational coupling constant. The action is well-defined only in regions of the manifold where the torsion scalars $\mathscr{V}$, $\mathscr{A}$, and $\mathscr{T}$ remain finite~\cite{lopez2025black, coley2025black}; this requirement plays a central role in our later analysis of the LH. By varying the action with respect to the tetrad we obtain the FE~\cite{lopez2025black}
\begin{equation}
W_{\mu\nu} = \kappa\, \Theta_{\mu\nu},
\end{equation}
where $\Theta_{\mu\nu}$ is the energy--momentum tensor. Since classical matter sources have a symmetric energy--momentum tensor, $\Theta_{[\mu\nu]}=0$, the AFE reduces to a constraint on the geometry, $W_{[\mu\nu]}=0$, which can equivalently be obtained by varying the action with respect to the spin connection~\cite{krvsvsak2019teleparallel}. This constraint takes the explicit form
\begin{dmath}\label{FEA}
W_{\al\mu\nu\ar} :-\frac{2}{3}\, \nabla^{\rho} \mathscr{V}_{[\mu} g_{\nu]\rho}-\frac{c_{a}}{3} \left(\frac{2}{3} \left(\epsilon_{\mu\nu\rho\gamma} \mathscr{T}_{\sigma}{}^{\rho\gamma} - 2\,\epsilon_{\sigma\rho\gamma[\nu} \mathscr{T}_{\mu]}{}^{\rho\gamma}\right)\mathscr{A}^{\sigma}+ \epsilon_{\mu\nu\sigma\rho} (\mathscr{V}^{\rho}\mathscr{A}^{\sigma}- \nabla^{\rho}\mathscr{A}^{\sigma})\right) + \frac{c_{t}}{2} \left(\mathscr{V}^{\rho}\mathscr{T}_{\rho\al\mu\nu\ar} - 2\, \nabla^{\rho}\mathscr{T}_{\rho\al\mu\nu\ar}\right)\mathrel{\nobreak=} 0  .
\end{dmath}
In the TEGR case, where $(c_{a}, c_{v}, c_{t}) = (3/2, -2/3, 2/3)$, this equation is identically satisfied. The SFE is then given by
\begin{dmath}\label{FES}
W_{\sl\mu\nu\sr} :-\frac{2}{3}\!\left(-\frac{1}{2} g_{\mu\nu}\mathscr{V}+ g_{\mu\nu}\nabla^{\rho}\mathscr{V}_{\rho} - \nabla^{\rho}\mathscr{V}_{\sl\mu} g_{\nu\sr\rho}\right)+ c_{a}\!\left(\frac{1}{6} g_{\mu\nu}\mathscr{A}+ \frac{1}{3} \mathscr{A}_{\mu}\mathscr{A}_{\nu} + \frac{4}{9} \epsilon_{\sigma\rho\gamma\sl\mu} \mathscr{T}_{\nu\sr}{}^{\rho\gamma} \mathscr{A}^{\sigma} \right) + c_{t}\!\left(\frac{1}{3} g_{\mu\nu} (\mathscr{T}-\mathscr{T}_{\alpha\sigma\rho}\mathscr{T}^{\alpha\rho\sigma})+ \frac{8}{3} \mathscr{T}^{\rho}{}_{\al\sigma\nu\ar} \mathscr{T}^{\sigma}{}_{\al\mu\rho\ar}- 2\, \epsilon_{\sigma\rho\alpha\sl\mu} \mathscr{T}_{\nu\sr}{}^{\rho\alpha} \mathscr{A}^{\sigma} \right. \left.+ \nabla^{\rho}(\mathscr{T}_{\mu\nu}{}{}^{\sigma}- \mathscr{T}^{\sigma}{}_{\sl\mu\nu\sr})- \frac{1}{2} \mathscr{V}^{\rho}(\mathscr{T}_{\mu\nu}{}{}^{\sigma}
 - \mathscr{T}^{\sigma}{}_{\sl\mu\nu\sr})\right)\mathrel{\nobreak=} \kappa\, \Theta_{\sl\mu\nu\sr} .
\end{dmath}

The Lagrangian in~\eqref{Lngr} reduces to a rescaled version of the TEGR Lagrangian for specific choices of the parameters \(c_a\), \(c_v\), and \(c_t\). Two distinct classes of models must therefore be considered, depending on whether \( c_v = 0 \) or \( c_v \neq 0 \)~\cite{lopez2025black}. When \( c_v = 0 \), TEGR cannot be recovered in any limit, and the Lagrangian reduces to a 2-parameter model that lies outside the family considered here. When \( c_v \neq 0 \) (the generic case), the Lagrangian can be normalized by dividing the entire expression by \(-3c_v/2\), which is equivalent to fixing \( c_v = -2/3 \). This yields the normalized form
\begin{equation}
\mathcal{L}
    = c_{a}\,\mathscr{A} + c_{t}\,\mathscr{T} - \frac{2}{3}\,\mathscr{V}
    = T + b_{1}\,\mathscr{T} - \frac{9}{4}\, b_{3}\,\mathscr{A} \, ,
\end{equation}
where we have used Eq.~\eqref{TS} and introduced the reparametrization
\begin{equation}\label{bp}
b_{1} = c_{t} - \frac{2}{3},
\qquad
b_{2} = 3c_{t},
\qquad
b_{3} = \frac{2}{3} - \frac{4c_{a}}{9} \, .
\end{equation}
These parameters originally depend on \( c_{v} \), as discussed in~\cite{lopez2025black}. In the normalized formulation of the theory, where \( c_{v} = -2/3 \), one immediately obtains the relation
\begin{equation}\label{norm}
b_{2} = 2 + 3 b_{1} .
\end{equation}
As a consequence, NGR depends only on the two remaining parameters \( c_{a} \) and \( c_{t} \), or equivalently on \( b_{1} \) and \( b_{3} \). This is the parametrization adopted here, as it allows the contributions controlled by these parameters to be interpreted naturally as deviations from the TEGR limit.

Using the parametrization~\eqref{bp}, together with the tetrad~\eqref{TetradSS} and the spin connection~\eqref{SpinV}, the nonzero components of the AFE~\eqref{FEA} and the SFE~\eqref{FES} can be written as follows
\begin{subequations}\label{FEV}
\begin{dmath}\label{FEA1}
W_{ \al tr \ar}  : \, \frac{b_{1}}{2} \left[(A_{1}r^{2}/A_{2})\psi'\right]' + A_{1}\left(b_{1}\cos\chi + b_{3}r[\cos\chi]' \right) \sinh\psi \mathrel{\nobreak=} 0   ,
\end{dmath}
\vskip -6pt
\begin{dmath}\label{FEA2}
W_{ \al \theta\phi \ar}  : \, \frac{1}{2}(b_{1}+b_{3}) \left[(A_{1}r^{2}/A_{2})\chi'\right]' - A_{1}\left(b_{3}r\,[\ln A_{1}]'-b_{1}+b_{3} \right)\cosh\psi \sin\chi
- b_{3}A_{1}\left([\cosh\psi]' + 2A_{2}\cos\chi \right)\sin\chi \mathrel{\nobreak=} 0   , 
\end{dmath}
\vskip -6pt
\begin{dmath}\label{FES1}
W^{t}{}_{t} : \, -\frac{F_{1}}{A_{2}{}^{2}} +\frac{1}{2}(b_{1}+b_{3})\left(\frac{\chi'}{A_{2}}\right)^{2}-\frac{b_{1}}{2}\left(\frac{\psi'}{A_{2}}\right)^{2}+\frac{2b_{3}}{r^2} \sin^{2}\chi \linebreak + \frac{2}{rA_{2}}\left( \frac{b_{1}}{r}\cos\chi+b_{3}[\cos\chi]'\right)\cosh\psi\mathrel{\nobreak=} \kappa \Theta^{t}{}_{t} \, ,
\end{dmath}
\vskip -6pt
\begin{dmath}\label{FES2}
W^{r}{}_{r} : \, \frac{F_{2}}{A_{2}{}^{2}}-\frac{1}{2}(b_{1}+b_{3})\left(\frac{\chi'}{A_{2}} \right)^{2} +\frac{b_{1}}{2}\left(\frac{\psi'}{A_{2}}\right)^{2} +\frac{2b_{3}}{r^2} \sin^{2}\chi \mathrel{\nobreak=}\kappa \Theta^{r}{}_{r} \, ,
\end{dmath}
\vskip -6pt
\begin{dmath}\label{FES3}
W^{\theta}{}_{\theta} : \, \frac{F_{3}}{A_{2}{}^{2}} +\frac{1}{2}(b_{1}+b_{3})\left(\frac{\chi'}{A_{2}}\right)^{2}-\frac{b_{1}}{2}\left(\frac{\psi'}{A_{2}} \right)^{2}-\frac{b_{1}}{r A_{2}}[\cosh\psi]'\cos\chi -\frac{1}{r A_{2}}\left(b_{1}[\ln A_{1}]'\cos\chi+(b_{1}-b_{3})[\cos\chi]' \right)\cosh\psi\mathrel{\nobreak=}\kappa \Theta^{\theta}{}_{\theta} \, ,  
\end{dmath}
\vskip -6pt
\begin{dmath}\label{FES12}
W_{ \sl tr \sr}  : \, -\frac{b_{1}}{2} \left[(A_{1}r^{2}/A_{2})\psi'\right]'- A_{1}\left(b_{1}\cos\chi +b_{3}r[\cos\chi]' \right)  \sinh\psi\mathrel{\nobreak=}0 \, .
\end{dmath}
\end{subequations}
Here we have introduced a set of functions \( F_{i} \), whose explicit expressions are given in Appendix~\eqref{FV}. We have also raised one index in the diagonal components of the SFE, (\ref{FES1}--\ref{FES3}), in order to remove the metric factors present in the energy--momentum tensor. This collects all purely geometric contributions on the left-hand side of the equations. Notably, Eqs.~\eqref{FEA1} and~\eqref{FES12} differ only by an overall sign, so that they are not independent: satisfying one automatically ensures the other, and consequently $W_{tr} = W_{[tr]} + W_{(tr)} = 0$~\cite{lopez2025black}. 

In addition, inspection of the full set of FE~\eqref{FEV} shows that the choice $\chi = n\pi$ decouples the parameter $b_{3}$ from the equations, while imposing $b_{1}=b_{3}=0$ simultaneously decouples both $\chi$ and $\psi$ from the system. Consequently, in the SSS case, TEGR can be recovered in two distinct ways: either by setting $b_{1}=b_{3}=0$, or by taking $b_{1}=0$ together with $\chi = n\pi$. This degeneracy means that, in the SSS sector, the effective parameter space of NGR collapses to TEGR along two inequivalent directions.

\subsection{NGR parameter space}\label{NGRps}

In general, NGR is characterized by two free parameters under an appropriate normalization, namely \( c_{t} \) and \( c_{a} \), or equivalently by \( b_{1} \) and \( b_{3} \) through the relations in~\eqref{bp}~\cite{lopez2025black}. These parameters are a priori arbitrary and are constrained only by global consistency requirements of the theory. One such requirement is the recovery of TEGR, which occurs for $c_{t} = 2/3$ and $c_{a} = 3/2$, or equivalently
\begin{equation}\label{TEGRlim}
b_{1} = b_{3} = 0.
\end{equation}
This condition is satisfied whenever the NGR model under consideration admits the limits \(b_{1}\to 0\) and \(b_{3}\to 0\) (with the latter being optional if $\chi=n\pi$, since in that case $b_{3}$ decouples from the FE). These limits ensure consistency with solar-system tests, including the Newtonian limit and its post-Newtonian (PPN) corrections~\cite{hayashi1979new, hayashi1981addendum, bahamonde2025revisiting}. We refer to NGR models satisfying these limits as models with an appropriate Newtonian limit.

However, additional physical requirements further restrict the parameter space. In particular, the existence of a propagating spin-2 mode and the absence of ghosts impose nontrivial constraints on \( b_{1} \) and \( b_{3} \). These conditions have been extensively analyzed in~\cite{bahamonde2025revisiting}, where a classification of NGR theories into 9 types is presented in terms of the three parameters $(c_{a}, c_{v}, c_{t})$. Re-expressing that classification in our parametrization and applying the normalization $b_{2} = 2 + 3 b_{1}$, the original parameter conditions translate into restricted conditions on $(b_{1}, b_{3})$ alone, yielding the reduced classification summarized in Table~\ref{GFC}.
\begin{table}[H]
\centering
\renewcommand{\arraystretch}{1.4}
\begin{tabular}{|c|c|c|}
\hline
\text{Theory} & \text{Parameter} & \text{Status / } \\
\noalign{\vskip -6pt}
\text{Type}   & \text{space}     & \text{Classification} \\
\hline
I   & Generic & Impossible (ghosts) \\
II  & $b_{1} = -\tfrac{2}{3}$ & DNPS-2 \\
III & $-\tfrac{2}{3} < b_{1} < 0$, $b_{3} = -b_{1}$ & Ghost-free \\
IV  & $b_{1} = 0$, $b_{3} > 0$ & Ghost-free \\
V   & $b_{1} = b_{3} = 0$ & TEGR \\
\hline
\end{tabular}
\caption{Classification of NGR theories obtained by translating the original 9-type classification of~\cite{bahamonde2025revisiting} into the normalized parametrization $b_{2}=2+3b_{1}$. ``DNPS-2'' indicates a model that does not propagate a spin-2 field.}
\label{GFC}
\end{table}
Let us now provide a brief description of the different types of NGR models listed in Table~\ref{GFC}:
 \begin{itemize}
\item[I:] This type cannot avoid ghost instabilities. 
\item[II:] This type does not propagate spin-2 particles (DNPS-2).
\item[III:] This type of models is ghost-free along the line \(b_{3}=-\,b_{1}\) within the parameter range $-2/3 < b_1 < 0$, and it admits a Newtonian limit, since both deviation parameters $b_{1}$ and $b_{3} = -b_{1}$ vanish as \(b_{1}\to 0^{-}\).
\item[IV:] This type is ghost-free whenever \(b_{3} > 0\) and admits a Newtonian limit either as $b_{3}\to 0^{+}$ or, in the SSS case considered here, automatically when $\chi = n\pi$ (which decouples $b_{3}$ from the FE).  
\item[V:] This type coincides with TEGR itself.
\end{itemize}

As a direct consequence of the normalization $b_{2}=2+3b_{1}$ and the ghost-free conditions of~\cite{bahamonde2025revisiting}, the physically viable NGR models all lie within the interval
\begin{equation}\label{bphy}
-\frac{2}{3} < b_{1} \leq 0 .
\end{equation}
Types~III, IV, and V satisfy all viability requirements, namely ghost-freedom, propagation of a spin-2 mode, and the existence of a Newtonian limit, and lie within the interval~\eqref{bphy}. Type~I is excluded because it cannot avoid ghost instabilities, while Type~II is excluded because it does not propagate a spin-2 mode. We will therefore restrict the subsequent analysis to the physically admissible Types~III, IV, and V.

\section{Vacuum black holes in NGR}\label{vacuum}

In this section we provide an overview of the detailed analysis presented in~\cite{lopez2025black}, where SSS vacuum configurations in NGR were systematically investigated; this overview establishes the framework that we will extend to the non-vacuum case in the following sections. Since the AFE do not depend on matter fields, the results of that analysis regarding the AFE are equally applicable to both vacuum and non-vacuum scenarios.

In~\cite{lopez2025black}, it was shown that obtaining exact solutions to the vacuum AFE and SFE~\eqref{FEV} is highly nontrivial when the functions \( \chi \) and \( \psi \) are treated as arbitrary. To explore whether NGR admits black hole geometries under these conditions, a perturbative method was developed, following the strategy employed in~\cite{coley2025black} for black holes in \( F(T) \) gravity. The analysis fixes a convenient coordinate gauge by choosing \( A_{3} = r \) and imposes the LH condition by requiring Eq.~\eqref{theAH} to vanish. This condition can be written as
\begin{equation}\label{localH}
\theta_{\sl \ell \sr} = \frac{\sqrt{2}}{r}\, a_{2}=0, 
\qquad 
a_{2} = \frac{1}{A_{2}} .
\end{equation}

Assuming that \( a_{2}(r_{h}) = 0 \), we introduce a perturbative parameter \( \epsilon \) and write the radial coordinate as \( r = r_{h} + \epsilon \) with \( \epsilon \to 0^{+} \). Under this assumption, we propose
\begin{equation}\label{lhan}
a_{2} = \epsilon^{p}(\alpha_{1} + \alpha_{2} \epsilon),
\end{equation}
with \( p > 0 \). Since \( a_{2} = 1/A_{2} \), and assuming a consistent perturbative structure for the remaining arbitrary functions, we adopt the following ansatz~\cite{coley2025black, lopez2025black}:
\begin{equation} \label{pertAXY}
A_{1} = \epsilon^{q} (\beta_{1} + \beta_{2} \epsilon), 
\qquad 
A_{2} = \frac{\epsilon^{-p}}{\alpha_{1} + \alpha_{2} \epsilon}, 
\qquad 
\chi = \epsilon^{u} (\chi_{0} + \gamma_{1} \epsilon), 
\qquad 
\psi = \epsilon^{v} (\psi_{0} + \gamma_{2} \epsilon),
\end{equation}
where \( q \), \( u \), and \( v \) are arbitrary constants. Using the ansatz~\eqref{pertAXY}, we rewrite the AFE~\eqref{FEA1} and~\eqref{FEA2} in terms of the perturbation parameter \( \epsilon \), retaining terms up to first order, to obtain:
\begin{subequations}\label{FEAp}
\begin{dmath}
W_{\al tr \ar} :\, -\frac{1}{2}b_{\on}\epsilon^{1+v}\gamma_{\tw}
\!\left(\frac{\alpha_{\tw}^{2}}{\alpha_{\on}^{2}}
+\frac{\beta_{\tw}^{2}}{\beta_{\on}^{2}}
+\frac{2}{r_{\h}^{2}}\right)
+\frac{1}{2}b_{\on}\epsilon^{-2+v}\psi_{\n}v(p+q+v-1)+G_{1}(\epsilon) \mathrel{\nobreak=}0 \, ,
\end{dmath}
\vskip -6pt
\begin{dmath}
W_{[\theta\phi]} : \, 
-(b_{\on}+b_{\tr})\,\epsilon^{1+u}\,\frac{\gamma_{\on}}{2}\!\left(\frac{\alpha_{\tw}^{2}}{\alpha_{\on}^{2}} + \frac{\beta_{\tw}^{2}}{\beta_{\on}^{2}} + \frac{2}{r_{\h}^{2}}\right)
+\frac{1}{2}(b_{\on}+b_{\tr})\,\chi_{\n}\,\epsilon^{-2+u}\,u(-1+p+q+u)+G_{2}(\epsilon)
\mathrel{\nobreak=}0.
\end{dmath}
\end{subequations}
Here we have introduced the functions \( G_{1}(\epsilon) \) and \( G_{2}(\epsilon) \), defined explicitly in Appendix~\eqref{G1} and~\eqref{G2}, respectively. We analyze the system of equations~\eqref{FEAp} for the nine possible combinations of the parameters \(u\) and \(v\), which can be grouped into the following categories~\cite{lopez2025black}:
\begin{center}
\begin{tabular}{l@{\hspace{1.5cm}}l@{\hspace{1.5cm}}l}
  1)\, $u > 0$ and $v > 0$ 
& 4)\, $u < 0$ and $v > 0$
& 7)\, $u = 0$ and $v > 0$ \\
  2)\, $u > 0$ and $v < 0$ 
& 5)\, $u < 0$ and $v < 0$ 
& 8)\, $u = 0$ and $v < 0$ \\
  3)\, $u > 0$ and $v = 0$ 
& 6)\, $u < 0$ and $v = 0$ 
& 9)\, $u = 0$ and $v = 0$
\end{tabular}
\end{center}

For each case, we examine whether the AFE are satisfied order by order, focusing on the leading contributions, i.e.~terms of the form \( \epsilon^{w} \) with \( w \leq 0 \). This procedure allows us to identify 55 solution branches, which are collected in~\cite{lopez2025black}. Similarly, we can express the SFE (\ref{FES1}--\ref{FES3}) in vacuum, that is \( \Theta^{\mu}{}_{\nu}=0 \), in terms of the perturbation parameter \( \epsilon \) using the ansatz~\eqref{pertAXY}. Retaining terms up to first order, and this time keeping both indices lowered to eliminate the extra factor of \( 1/A_{2}{}^{2} \), we obtain~\cite{lopez2025black}:
\begin{subequations}\label{FESp}
\begin{dmath}
W_{tt} : \frac{b_{\on} q (-2 + 2p + q)}{2 \epsilon^{2}}
+ \frac{-\alpha_{\on} (2 + 3 b_{\on}) \beta_{\on} p
+ \alpha_{\tw} b_{\on} \beta_{\on} q r_{\h}
+ \alpha_{\on} b_{\on} (p + q) (2 \beta_{\on} + \beta_{\tw} r_{\h})}{\alpha_{\on} \beta_{\on} \epsilon r_{\h}} +G_{3}(\epsilon)\mathrel{\nobreak=}0,
\end{dmath}
\vskip -12pt
\begin{dmath}
W_{rr} : \, -\frac{b_{\on} q^{2}}{2 \epsilon^{2}}
+\frac{q\,\big[(2+3b_{\on})\beta_{\on}-b_{\on}(2\beta_{\on}+\beta_{\tw}r_{\h})\big]}{\beta_{\on}\,\epsilon\,r_{\h}}+G_{4}(\epsilon)\mathrel{\nobreak=}0,
\end{dmath}
\vskip -12pt
\begin{dmath}
W_{\theta\theta} : \,
\frac{q\big[(2+3b_{\on})(-1+p+q)-b_{\on}(-2+2p+q)\big]}{2\epsilon^{2}}
+\frac{\alpha_{\tw}(2+b_{\on})\beta_{\on}q
+\alpha_{\on}\beta_{\tw}\big[(2+b_{\on})p+4q\big]}{2\alpha_{\on}\beta_{\on}\epsilon}
+\frac{(2-b_{\on})\beta_{\on}(p+q) 
+4b_{\on}\beta_{\tw}q\,r_{\h}}{2\beta_{\on}\epsilon\,r_{\h}} +G_{5}(\epsilon)\mathrel{\nobreak=}0 ,
\end{dmath}
\end{subequations}
where $G_{3}(\epsilon)$, $G_{4}(\epsilon)$, and $G_{5}(\epsilon)$ are defined in~\eqref{G3},~\eqref{G4}, and~\eqref{G5}, respectively (see Appendix~\ref{AuxF}). Using parameter values that satisfy the AFE, we analyzed the corresponding SFE, as reported in~\cite{lopez2025black}. A detailed comparison revealed significant overlap among these branches, with several cases related by parameter identifications or by one branch representing a more general form of another. 

Substituting the parameter values of each of the 55 AFE branches into the SFE~\eqref{FESp} yields 55 corresponding sets of equations. By performing a detailed comparison of each resulting set with all others, we identify which AFE branches produce identical SFE. The branches within a given class yield identical SFE at leading perturbative order and therefore represent the same physical solution up to reparametrization. This allows us to group the AFE branches into the nine classes shown in Table~\ref{ClassTable}, where we have kept only the most general branch within each class; of these, only class~A admits solutions that are not trivially equivalent to TEGR (class~T).
\begin{table}[!h]
\centering
\renewcommand{\arraystretch}{1.15}
\begin{tabular}{|c|c|c|c|c|c|c|c|c|c|c|c|c|c|c|}
\hline
\multicolumn{3}{|c|}{NGR} &
\multicolumn{3}{c|}{$\chi$} &
\multicolumn{3}{c|}{$\psi$} &
\multicolumn{3}{c|}{$A_{1}$} &
\multicolumn{3}{c|}{$A_{2}$} \\
\hline
\text{Class} & $b_{\on}$ & $b_{\tr}$ &
$u$ & $\chi_{0}$ & $\gamma_{\on}$ &
$v$ & $\psi_{0}$ & $\gamma_{\tw}$ &
$q$ & $\beta_{\on}$ & $\beta_{\tw}$ &
$p$ & $\alpha_{\on}$ & $\alpha_{\tw}$ \\
\hline
T & $0$ & $0$ & $0$ & $\;$ & $\;$ & $0$ & $\;$ & $\;$ & $\;$ & $\;$ & $\;$ & $(0,\infty)$ & $\;$ & $\;$ \\
\hline
A & $\;$ & $\;$ & $0$ & $n\pi$ & $0$ & $0$ & $0$ & $0$ & $\;$ & $\;$ & $\;$ & $(0,\infty)$ & $\;$ & $\;$ \\
\hline
B & $0$ & $\;$ & $(0,1]$ & $0$ & $\;$ & $0$ & $\;$ & $\;$ & $-p-u$ & $\;$ & $\;$ & $(0,\infty)$ & $\;$ & $\;$ \\
\hline
C & $\;$ & $-b_{\on}$ & $0$ & $n\pi$ & $\;$ & $0$ & $0$ & $0$ & $0$ & $\;$ & $\;$ & $(0,\tfrac{1}{2})$ & $\;$ & $\;$ \\
\hline
D & $\;$ & $\;$ & $0$ & $n\pi$ & $0$ & $(0,1]$ & $0$ & $\;$ & $-p-v$ & $\;$ & $\;$ & $(0,\infty)$ & $\;$ & $\;$ \\
\hline
E & $\;$ & $0$ & $(0,1]$ & $\;$ & $0$ & $0$ & $0$ & $0$ & $1-p-u$ & $\;$ & $-\frac{2\beta_{\on}}{r_{h}}$ & $(0,\infty)$ & $\;$ & $0$ \\
\hline
F & $\;$ & $\;$ & $(0,1]$ & $0$ & $0$ & $(0,1]$ & $\;$ & $0$ & $1-p-v$ & $\;$ & $-\frac{2\beta_{\on}}{r_{h}}$ & $(0,\infty)$ & $\;$ & $0$ \\
\hline
G & $\;$ & $\;$ & $(0,1]$ & $0$ & $\;$ & $0$ & $0$ & $0$ & $-p-u$ & $\;$ & $\;$ & $(0,\infty)$ & $\;$ & $\;$ \\
\hline
H & $\;$ & $\;$ & $0$ & $n\pi$ & $0$ & $0$ & $0$ & $\;$ & $-p$ & $\;$ & $a\beta_{\on}$ & $(0,\infty)$ & $\;$ & $d\alpha_{\on}$ \\
\hline
\end{tabular}
\caption{Classification of the 55 AFE branches by equivalence under the SFE. Branches within the same class yield identical SFE at leading perturbative order; only the most general branch of each class is shown. Class~T corresponds to TEGR. Classes B--H are either particular cases of class~A or trivial, and so do not contribute physically distinct solutions beyond those found in class~A. Blank entries indicate unconstrained parameters. The constant $d$ in class~H is defined by $d = -\left(a + 2/r_{\h}\right)$, with $a$ an arbitrary constant.}
\label{ClassTable}
\end{table}

Within this classification, and excluding the TEGR cases, admissible solution branches (i.e., branches that satisfy both AFE and SFE at leading perturbative order without reducing to $b_{1}=b_{2}=b_{3}=0$) are found only in class~A (see Table~\ref{SolTS})~\cite{lopez2025black}. Imposing the parameter values that solve both the AFE and SFE at leading perturbative order, we find that the choices \(\chi=n\pi\) and \(\psi=0\) are necessary for SSS NGR geometries in vacuum using the gauge \(A_{3}=r\). Using these conditions together with the perturbative ansatz~\eqref{pertAXY}, we then evaluate the torsion scalars~\eqref{TScalars} and find that all remain finite in the limit \(\epsilon\to 0^{+}\), as summarized in Table~\ref{SolTS}.
\begin{table}[!h]
\centering
\renewcommand{\arraystretch}{1}
\begin{tabular}{|c|c|c|c|c|c|c|c|c|c|c|c|c|c|}
\hline
\multicolumn{4}{|c|}{\text{NGR}} & \multicolumn{3}{c|}{$A_1$} & \multicolumn{3}{c|}{$A_2$} & \multicolumn{3}{c|}{$\epsilon \to 0^{+}$} \\
\cline{1-13}
\# & $b_{\on}$   & $b_{\tr}$ & Lagrangian & $q$ & $\beta_{\on}$ & $\beta_{\tw}$ & $p$ & $\alpha_{\on}$ & $\alpha_{\tw}$   & $\mathscr{T}$ & $\mathscr{V}$ & $\mathscr{A}$  \\
\hline
1 & $-\frac{2}{3}$   &  & $c_{a} \mathscr{A} - \frac{2}{3} \mathscr{V}$ & $-\frac{2\delta}{\alpha_{\on} r_{\h}}$ &  & $-\frac{2\beta_{\on}}{r_{\h}}$ & $1$ &  & $-\frac{\alpha_{\on}}{r_{\h}}$ & $\frac{9}{r_{\h}^2}$ & $0$ & $0$  \\[11pt]
2.a & $2$  & & $c_{a} \mathscr{A} + \frac{8}{3} \mathscr{T} - \frac{2}{3} \mathscr{V}$ & $0$ &  & $0$ & $1$ & $-\frac{\delta}{r_{\h}}$ &  & $\frac{1}{r_{\h}^2}$ & $\frac{4}{r_{\h}^2}$ & $0$  \\[11pt]
2.b & $2$   &  & $c_{a} \mathscr{A} + \frac{8}{3} \mathscr{T} - \frac{2}{3} \mathscr{V}$ & $0$ &  & $\frac{4\beta_{\on}}{r_{\h}}$ & $1$ & $\frac{\delta}{r_{\h}}$ &  & $\frac{1}{r_{\h}^2}$ & $\frac{4}{r_{\h}^2}$ & $0$ 
 \\[11pt]
\hline
\end{tabular}
\caption{Parameter values satisfying both the AFE and SFE, and the behavior of the torsion scalars as \( \epsilon \to 0^{+} \), using \( \chi = n\pi \), \( \psi = 0 \) and $\delta=\pm1$. Blank entries indicate unconstrained parameters. Adapted from~\cite{lopez2025black}.}
\label{SolTS}
\end{table}

The information in Table~\ref{SolTS} shows that case~1 corresponds to a one-parameter NGR model at the Lagrangian level. However, since \( \chi = n\pi \) decouples \( b_{3} \) from the FE, the model effectively reduces to a zero-parameter theory. In this case, \( b_{1} = -2/3 \), which classifies the model as Type~II in Table~\ref{GFC}. This class does not support a propagating spin-2 field and therefore admits no gravitational waves. However, the geometry remains regular at the LH, and the region \( r = r_{\h} \) and its interior (excluding \( r = 0 \)) form part of the manifold.

Cases~2.a and~2.b represent two branches of the same type and likewise describe a one-parameter model at the Lagrangian level. As before, the choice \( \chi = n\pi \) eliminates the dependence on \( b_{3} \) in the FE, reducing the theory to a zero-parameter model with \( b_{1} = 2 \). Since \( b_{1} = 2 \) lies outside the physically admissible region identified in~\eqref{bphy}, this corresponds to Type~I in Table~\ref{GFC} and the model inevitably exhibits ghost instabilities~\cite{bahamonde2025revisiting, lopez2025black}. Nevertheless, the geometry is regular at the LH, and the region \( r = r_{\h} \) and its interior (excluding \( r = 0 \)) remain admissible parts of the manifold.

This analysis shows that the vacuum models in Table~\ref{SolTS} are, in principle, well-behaved at the LH. However, such models exhibit important unphysical features, and so no further investigation was carried out; for instance, the lower-order conditions from the AFE and SFE were not explicitly considered~\cite{lopez2025black}. We therefore conclude that NGR is unable to describe vacuum black hole configurations while maintaining physical consistency with key requirements such as the Newtonian limit, ghost stability, and propagating spin-2 modes. In the next section, we extend this analysis to the non-vacuum case.

\section{Non-vacuum black holes in NGR}\label{AFEp9}

In NGR, the search for SSS vacuum black hole geometries forces the free parameters of the theory to take specific values. These values coincide with regions of parameter space corresponding to known pathological models — namely, those that either lack a propagating spin-2 mode, unavoidably contain ghost instabilities, or lack a Newtonian limit~\cite{lopez2025black}. It is important to emphasize that the mechanism fixing \( b_{1} \) to such unphysical values arises specifically and exclusively from the SSS vacuum SFE (i.e., \( W_{\mu\nu} = 0 \)). In the presence of matter, however, the SFE become inhomogeneous, \( W_{\mu\nu} = \kappa\,\Theta_{\mu\nu} \), and the algebraic constraints responsible for fixing \( b_{1} \) may no longer apply. The matter source therefore plays a role analogous to that of, for instance, a stellar interior in GR — but with an even sharper effect, since here it can directly modify the algebraic constraints on the parameter space rather than merely the geometric configuration. Thus, although our vacuum analysis shows that SSS vacuum configurations in NGR are only realized at unphysical points in parameter space, it remains an open question whether SSS non-vacuum configurations admit black-hole-like solutions within the physically viable region of parameter space identified in~\eqref{bphy}.

Our approach to the SSS non-vacuum case follows the perturbative method previously developed in~\cite{lopez2025black}. As before, we assume the existence of a LH~\eqref{localH}, which motivates the perturbative ansatz~\eqref{pertAXY} and allows the geometrical sector of the FE to be expressed in terms of the parameter \( \epsilon \). For consistency, the matter sector must also be expanded within the same perturbative framework, and the resulting matter quantities are constrained by the conservation equation $D^{\mu}\Theta_{\mu\nu}=0$ (where $D_{\mu}$ denotes the covariant derivative with respect to the Levi-Civita connection), which links the components of the energy--momentum tensor to the underlying geometry. As the matter source we adopt a comoving perfect fluid coupled with an electromagnetic field, treated as two separately conserved sectors. This choice provides the simplest physically transparent setup compatible with SSS configurations while allowing for the most general charged black-hole-like solutions.

\subsection{Energy-momentum conservation}

We now make the matter sector specified in the previous paragraph explicit. The total 
energy--momentum tensor decomposes as
\begin{equation}\label{EMT}
    \Theta^{\mu}{}_{\nu} \;=\; \Theta_{\text{\tiny(F)}}^{\mu}{}_{\nu}\;+\; \Theta_{\text{\tiny(E)}}^{\mu}{}_{\nu},
\end{equation}
where the perfect-fluid and electromagnetic contributions take the standard 
forms~\cite{wald2010general}
\begin{equation}
    \Theta_{\text{\tiny(F)}}^{\mu}{}_{\nu}= (\rho + P)\, u^{\mu} u_{\nu}+ P\, \delta^{\mu}_{\nu},
    \qquad
    \Theta_{\text{\tiny(E)}}^{\mu}{}_{\nu}= F^{\mu\alpha} F_{\nu\alpha}- \frac{1}{4}\,\delta^{\mu}_{\nu}F^{\alpha\beta} F_{\alpha\beta}.
\end{equation}
Here $u^{\mu}$ denotes the fluid four-velocity, and $F^{\mu\nu}$ is the electromagnetic field-strength tensor. The assumption of separate conservation corresponds to a neutral perfect fluid that does not interact directly with the electromagnetic field. In a SSS configuration, the only nonvanishing component of the field-strength tensor corresponds to a purely radial electric field, with coordinate components $F_{tr}=E(r)\,A_{1}A_{2}$. Solving the source-free Maxwell equation $\partial_{\mu}(\sqrt{-g}\,F^{\mu\nu})=0$ for its only nontrivial component yields
\begin{equation}\label{Ele}
    E(r)=\frac{Q_{0}}{r^{2}},
\end{equation}
where $Q_{0}$ is an integration constant interpreted as the conserved electric charge. This automatically guarantees the conservation of the electromagnetic energy--momentum tensor $\Theta_{\text{\tiny(E)}}^{\mu}{}_{\nu}$. In contrast, the conservation of the fluid contribution $\Theta_{\text{\tiny(F)}}^{\mu}{}_{\nu}$ is nontrivial and requires a dedicated treatment within the perturbative framework.

Considering that the left-hand side of the SFE (\ref{FES1}--\ref{FES3}) is expressed in terms of the perturbative parameter $\epsilon$, for consistency we require that the matter fields entering the energy--momentum tensor~\eqref{EMT} on the right-hand side adopt a structure analogous to~\eqref{pertAXY}. Accordingly, the pressure $P$ and energy density $\rho$ are expanded as
\begin{equation}\label{perPr}
P = \epsilon^{y}\!\left( P_{0} + \epsilon\, P_{1} \right), \qquad \rho = \epsilon^{z}\!\left( \rho_{0} + \epsilon\, \rho_{1} \right),
\end{equation}
where $y$ and $z$ are arbitrary real exponents, allowing the matter source to either vanish or diverge at the LH. Since the fluid and electromagnetic contributions are separately conserved, the conservation equation for the comoving perfect fluid takes the SSS TOV-like form~\cite{wald2010general, poisson2004relativist}
\begin{equation}\label{conv}
D_{\mu}\,\Theta_{\text{\tiny(F)}}^{\mu}{}_{\nu} = 0 
\quad\longrightarrow\quad
\frac{(P+\rho)\,A_{1}'}{A_{1}} + P' = 0 .
\end{equation}
Substituting the perturbative expressions~\eqref{perPr} for $P$ and $\rho$, and retaining terms up to first order in $\epsilon$, the conservation equation~\eqref{conv} becomes
\begin{dmath}\label{conEq}
\epsilon^{-1+y} P_{0} (q+y)+\frac{\epsilon^{\,y}\left(\beta_{\tw} P_{0} + \beta_{\on}P_{1}(1+q+y)\right)}{\beta_{\on}}-\frac{\beta_{\tw}\, \epsilon^{\,1+y}\!\left(\beta_{\tw} P_{0} - \beta_{\on} P_{1}\right)}{\beta_{\on}^{2}}
\\[4pt]+\epsilon^{-1+z} q\, \rho_{0}+\frac{\epsilon^{\,z}\left(\beta_{\tw}\rho_{0} + \beta_{\on} q\,\rho_{1}\right)}{\beta_{\on}} -
\frac{\beta_{\tw}\, \epsilon^{\,1+z}\!\left(\beta_{\tw}\rho_{0} - \beta_{\on}\rho_{1}\right)}{\beta_{\on}^{2}}
= 0 .
\end{dmath}
For the matter sector to contribute at the same perturbative order as the dominant geometric terms in the SFE, we restrict the exponents to $y \leq 1$ and $z \leq 1$. Otherwise, the matter contributions appear only at subleading orders and play no role in the analysis near the LH.

A priori, nothing in our setup imposes a fixed relation between the perturbative orders of $P$ and $\rho$. In cosmological settings a linear equation of state would immediately enforce $z = y$~\cite{coley2024spherically, vandenhoogen2023bianchi}, but for non-vacuum black holes no such relation is imposed. Instead, the behaviour of $P$ and $\rho$ is determined by the SFE (\ref{FES1}--\ref{FES3}). Even without an explicit equation of state, the structure of the SFE (\ref{FES1}--\ref{FES3}) naturally ties the perturbative orders of $P$ and $\rho$. The fluid enters only through the combinations $P$, $\rho$, and $\rho + P$, all of which appear at the same perturbative level when compared with the geometric sector (i.e., the left-hand side of Eqs.~\eqref{FESp}). If $y \neq z$, one of these quantities would dominate as $\epsilon \to 0^{+}$ while the other would be subleading; the conservation equation~\eqref{conEq}, together with the SFE~(\ref{FES1}--\ref{FES3}), would then generically force the subleading quantity to vanish at the dominant order — yielding inconsistent or trivial branches.

To ensure that the matter acts as a single, self-consistent perturbative source, we therefore impose
\begin{equation}\label{zy}
z = y .
\end{equation}
This choice guarantees that $P$ and $\rho$ contribute at the same perturbative order, keeping the combinations $P$, $\rho$, and $\rho + P$ balanced and ensuring a consistent impact on the geometry throughout the perturbative analysis. Then, using~\eqref{zy} and restricting to the nontrivial regime $y \leq 1$ within the perturbative framework, the conservation equation~\eqref{conEq} reduces to
\begin{dmath}\label{conper}
\epsilon^{-1+y}\bigl(P_{0}(q+y) + q\,\rho_{0}\bigr)
\;+\;
\frac{\epsilon^{y}\bigl(\beta_{\on} P_{1}(1+q+y) + \beta_{\tw}(P_{0}+\rho_{0}) + \beta_{\on} q\,\rho_{1}\bigr)}{\beta_{\on}}
\;+\;
\frac{\beta_{\tw}\,\epsilon^{1+y}\bigl(-\beta_{\tw}(P_{0}+\rho_{0}) + \beta_{\on}(P_{1}+\rho_{1})\bigr)}{\beta_{\on}^{2}}
= 0 \,.
\end{dmath}
Given the structure of Eq.~\eqref{conper}, three relevant regions $R_{1}$, $R_{2}$ and $R_{3}$ of analysis must be distinguished:
\begin{equation}
R_{1}: 0 < y \leq 1, 
\qquad
R_{2}: -1 < y \leq 0,
\qquad
R_{3}: y \leq -1 .
\end{equation}
These regions correspond to physically distinct behaviours of the matter source at the LH: in $R_{1}$, both pressure and energy density vanish at the LH ($P,\rho \to 0$ as $\epsilon \to 0^{+}$); in $R_{2}$, they are finite or mildly divergent, with the boundary case $y=0$ corresponding to constant leading values; and in $R_{3}$, both pressure and energy density strongly diverge at the LH. We now analyze each region in turn.

Focusing first on $R_{1}$, the leading-order contribution arises from the term proportional to $\epsilon^{-1+y}$, with all remaining terms being negligible in this regime. Consequently, for the conservation equation to hold at leading order, the admissible solution branches must correspond to the parameter values listed in Table~\ref{R1}.
\begin{table}[!h]
\centering
\renewcommand{\arraystretch}{1.3}
\begin{tabular}{|c|c|c|c|c|c|c|c|c|}
\hline
$R_{1}$ &
\multicolumn{3}{c|}{$A_{1}$} &
\multicolumn{3}{c|}{$P$} &
\multicolumn{2}{c|}{$\rho$} \\
\cline{1-9}
\# &
$q$ & $\beta_{\on}$ & $\beta_{\tw}$ &
$y$ & $P_{0}$ & $P_{1}$ &
$\rho_{0}$ & $\rho_{1}$ \\
\hline
1 &
 &  &  &
$(0,1]$ &
 &  &
$-\frac{P_{0}(q+y)}{q}$ &
 \\
\hline
2 &
$0$ &  &  &
$(0,1]$ &
$0$ &  &
 &  \\
\hline
\end{tabular}
\caption{Parameter values satisfying the conservation equation for $R_{1}$. Blank entries indicate unconstrained parameters. }
\label{R1}
\end{table}

In $R_{2}$, the leading-order contribution arises from the term proportional to $\epsilon^{-1+y}$, followed by the next-to-leading term of order $\epsilon^{y}$, with the remaining contribution being negligible in this regime. Consequently, for the conservation equation to be satisfied at leading order, the admissible solution branches must correspond to the parameter values listed in Table~\ref{R2}.

\begin{table}[!h]
\centering
\renewcommand{\arraystretch}{1.3}
\begin{tabular}{|c|c|c|c|c|c|c|c|c|}
\hline
$R_{2}$ &
\multicolumn{3}{c|}{$A_{1}$} &
\multicolumn{3}{c|}{$P$} &
\multicolumn{2}{c|}{$\rho$} \\
\cline{1-9}
\# &
$q$ & $\beta_{\on}$ & $\beta_{\tw}$ &
$y$ & $P_{0}$ & $P_{1}$ &
$\rho_{0}$ & $\rho_{1}$ \\
\hline
1 &
 &  &  &
$(-1,0]$ &
 &  &
$-\frac{P_{0}(q+y)}{q}$ &
$\frac{\beta_{\tw}P_{0}y-\beta_{\on}P_{1}q(1+q+y)}{\beta_{\on}q^{2}}$ \\
\hline
2 &
$0$ &  &  &
$(-1,0]$ &
$0$ &  &
$-\frac{\beta_{\on}P_{1}(1+y)}{\beta_{\tw}}$ &
 \\
\hline
3 &
$0$ &  &  &
$0$ &
 &  &
$-\frac{\beta_{\tw}P_{0}+\beta_{\on}P_{1}}{\beta_{\tw}}$ &
 \\
\hline
4 &
$0$ &  & $0$ &
$0$ &
 & $0$ &
 &
 \\
\hline
\end{tabular}
\caption{Parameter values satisfying the conservation equation for $R_{2}$. Blank entries indicate unconstrained parameters.}
\label{R2}
\end{table}

Finally, in $R_{3}$, the leading-order contribution again arises from the term proportional to $\epsilon^{-1+y}$, while all remaining terms appear at the next-to-leading orders. Thus, in this regime all contributions are relevant. Consequently, for the conservation equation to hold at leading order, the admissible solution branches are those listed in Table~\ref{R3}.
\begin{table}[!h]
\centering
\renewcommand{\arraystretch}{1.3}
\begin{tabular}{|c|c|c|c|c|c|c|c|c|}
\hline
$R_{3}$ &
\multicolumn{3}{c|}{$A_{1}$} &
\multicolumn{3}{c|}{$P$} &
\multicolumn{2}{c|}{$\rho$} \\
\cline{1-9}
\# &
$q$ & $\beta_{\on}$ & $\beta_{\tw}$ &
$y$ & $P_{0}$ & $P_{1}$ &
$\rho_{0}$ & $\rho_{1}$ \\
\hline
1 &
 &  & $0$ &
$(-\infty,-1]$ &
 &  &
$-\frac{P_{0}(q+y)}{q}$ &
$-\frac{P_{1}(1+q+y)}{q}$ \\
\hline
2 &
 &  &  &
$(-\infty,-1]$ &
$\frac{\beta_{\on}P_{1}q(1+y)}{\beta_{\tw}y+\beta_{\tw}qy}$ &
 &
$-\frac{\beta_{\on}P_{1}(1+y)(q+y)}{\beta_{\tw}(1+q)y}$ &
$-\frac{P_{1}(2+q+y)}{1+q}$ \\
\hline
3 &
 &  & $0$ &
$(-\infty,-1]$ &
 & $0$ &
$-\frac{P_{0}(q+y)}{q}$ &
$0$ \\
\hline
4 &
$-1$ &  &  &
$(-\infty,-1]$ &
 & $0$ &
$P_{0}(-1+y)$ &
$\frac{\beta_{\tw}P_{0}y}{\beta_{\on}}$ \\
\hline
5 &
$0$ &  & $0$ &
$-1$ &
$0$ &  &
 &  \\
\hline
\end{tabular}
\caption{Parameter values satisfying the conservation equation for $R_{3}$. Blank entries indicate unconstrained parameters.}
\label{R3}
\end{table}

The parameter values listed in Tables~\ref{R1}--\ref{R3} for the various ranges of $y$, together with the expression for the electric field in~\eqref{Ele}, ensure the conservation of the energy--momentum tensor in~\eqref{EMT}. These results provide the necessary input for the right-hand side of the SFE~(\ref{FES1}--\ref{FES3}); in the next subsection we combine these matter-conservation branches with the SFE to identify the full set of admissible SSS non-vacuum NGR configurations.

\subsection{Analysis of the SFE}

A preliminary examination of the SFE, applied to all branches that satisfy the AFE at leading order, leads to the equivalence classes listed in Table~\ref{ClassTable} (introduced in Section~\ref{vacuum}). We analyze the SFE for all classes in that table, excluding the TEGR class (first row). Building on the preceding results, we perform a perturbative analysis of the SFE by rewriting them using the ansatz~\eqref{pertAXY} and~\eqref{perPr}, together with the relation~\eqref{zy}, and retaining terms up to first order in the perturbative parameter $\epsilon$.

We then assess whether the SFE can be consistently satisfied within this framework. Since the constants $b_{1}$ and $b_{3}$ characterize the NGR models under consideration, we seek solutions that do not require fixing their values unless unavoidable. Throughout the analysis, we systematically use the information summarized in Tables~\ref{R1}--\ref{R3}, which ensures conservation of the energy--momentum tensor~\eqref{EMT}. The conservation equations are identically satisfied when $P_{0}=P_{1}=0$, $\rho_{0}=\rho_{1}=0$, and $Q_{0}=0$, corresponding to the pure vacuum sector previously analyzed in~\cite{lopez2025black}, reviewed in Sec.~\ref{vacuum} and shown to be pathological. This branch is therefore excluded from the present analysis.

Since the perturbative analysis of the SFE is systematic and repetitive, we present the full analysis only for class~A, which serves as a representative example. The remaining classes follow the same procedure; further details are provided in~\cite{lopez2026thesis}.

\subsubsection{Class A}

We begin our analysis of the SFE with the first class of interest, namely class~A, characterized by the parameter values in the second row of Table~\ref{ClassTable}. Using these parameter values, and retaining only the leading-order contributions, the SFE~(\ref{FES1}--\ref{FES3}) with perfect-fluid and electric-charge contributions can be written, in terms of the perturbative parameter $\epsilon$, as follows:
\begin{subequations}\label{SFEAp}
\begin{dmath} W^{t}{}_{t} : -\frac{1}{2}\alpha_{\on}^{2}b_{\on}\epsilon^{-2+2p}q(-2+2p+q) + \frac{-2+b_{\on}}{2r_{\h}^{2}} + \frac{\alpha_{\on}\epsilon^{-1+2p}\left[\alpha_{\on}\beta_{\on}\bigl((2+b_{\on})p-2b_{\on}q\bigr) - b_{\on}\bigl(\alpha_{\on}\beta_{\tw}(p+q) + \alpha_{\tw}\beta_{\on}q(-1+2p+q)\bigr)r_{\h}\right]}{\beta_{\on}r_{\h}} = -\frac{Q_{0}^{2}\kappa}{8\pi r_{\h}^{4}} - \epsilon^{y}\kappa\rho_{0} - \epsilon^{1+y}\kappa\rho_{1} \, ,
\end{dmath}
\vskip -12pt
\begin{dmath} W^{r}{}_{r} : -\frac{1}{2}\alpha_{\on}^{2}b_{\on}\epsilon^{-2+2p}q^{2} + \frac{-2+b_{\on}}{2r_{\h}^{2}} + \frac{\alpha_{\on}\epsilon^{-1+2p}q\left[\alpha_{\on}(2+b_{\on})\beta_{\on} - b_{\on}(\alpha_{\on}\beta_{\tw}+\alpha_{\tw}\beta_{\on}q)r_{\h}\right]}{\beta_{\on}r_{\h}} = \epsilon^{y}P_{0}\kappa + \epsilon^{1+y}P_{1}\kappa - \frac{Q_{0}^{2}\kappa}{8\pi r_{\h}^{4}} \, ,
\end{dmath}
\vskip -12pt
\begin{dmath} W^{\theta}{}_{\theta} : \frac{1}{2}\alpha_{\on}^{2}\epsilon^{-2+2p}q\bigl[(2+b_{\on})(-1+p)+2(1+b_{\on})q\bigr] + \frac{\alpha_{\on}^{2}\beta_{\tw}\epsilon^{-1+2p}\bigl[(2+b_{\on})p+4(1+b_{\on})q\bigr]}{2\beta_{\on}} + \frac{1}{2}\alpha_{\on}\alpha_{\tw}\epsilon^{-1+2p}q\bigl[(2+b_{\on})(-1+2p)+4(1+b_{\on})q\bigr] - \frac{(-1)^{n}\alpha_{\on}b_{\on}\epsilon^{-1+p}q}{r_{\h}} \linebreak - \frac{\alpha_{\on}^{2}(-2+b_{\on})\epsilon^{-1+2p}(p+q)}{2r_{\h}} \mathrel{\nobreak=} \epsilon^{y}P_{0}\kappa + \epsilon^{1+y}P_{1}\kappa + \frac{Q_{0}^{2}\kappa}{8\pi r_{\h}^{4}} \, .
\end{dmath}
\end{subequations}
We now analyze these equations order by order, assuming $p>0$ and using Tables~\ref{R1}--\ref{R3} to guide the search for branches that satisfy the SFE. This procedure leads to the results summarized in Table~\ref{AT}, where we have introduced the functions $g_{i}$ as defined in Appendix~\ref{gfun}. 

Note that since $\chi = n\pi$ in this class, the parameter $b_{3}$ decouples from the FE, as is evident from~\eqref{SFEAp}, and the parameter $b_{1}$ remains unrestricted. Moreover, several branches turn out to be qualitatively equivalent in Table~\ref{AT}: specifically, A.5 is equivalent to A.1, A.6 to A.2, and A.7 to A.4. This can be verified by direct comparison of the parameter values, which differ only by a shift of the matter exponent $y$ with a corresponding rescaling of the matter amplitudes; the physical content — the leading behaviour of $P$, $\rho$, and the geometry at the LH — is identical. Consequently, only four independent branches remain, namely A.1, A.2, A.3, and A.4.

Applying this procedure to classes B through H in Table~\ref{ClassTable} yields a total of 32 distinct branches satisfying the SFE, including the 7 branches arising from class~A discussed above. Although these results are explicitly presented in~\cite{lopez2026thesis}, they can also be independently reproduced using the information provided here.

\begin{table}[!h]
\centering
\renewcommand{\arraystretch}{1.2}
\begin{tabular}{|c|c|c|c|c|c|c|c|c|c|c|c|c|c|}
\hline
NGR &
\multicolumn{3}{c|}{$A_{1}$} &
\multicolumn{3}{c|}{$A_{2}$} &
\multicolumn{3}{c|}{$P$} &
\multicolumn{2}{c|}{$\rho$} &
$E$ &
CE \\
\hline
\# & $q$ & $b_{1}$ & $b_{2}$ & $p$ & $\alpha_{1}$ & $\alpha_{2}$ &
$y$ & $P_{0}$ & $P_{1}$ & $\rho_{0}$ & $\rho_{1}$ & $Q_{0}$ & \# \\
\hline
A.1 & & & & $(0,\infty)$ & 0 & & 0 &
$\frac{-2+b_{1}}{4 r_{h}^{2}\kappa}$ & &
$\frac{2-b_{1}}{4 r_{h}^{2}\kappa}$ &
$-\frac{P_{1}(1+q)}{q}$ & $g_{1}$ & $R_{2}$.1 \\
A.2 & & & & 1 & $g_{2}$ & & 0 & $g_{3}$ & &
$-g_{3}$ & $-\frac{P_{1}(1+q)}{q}$ & & $R_{2}$.1  \\
A.3 & 0 & & $g_{4}$ & $\frac12$ & & & 0 & $g_{5}$ & $g_{6}$ &
$g_{7}$ & & & $R_{2}$.3 \\
A.4 & 0 & & 0 & $(\frac12,\infty)$ & & & 0 &
$\frac{-2+b_{1}}{4 r_{h}^{2}\kappa}$ & 0 &
$\frac{2-b_{1}}{4 r_{h}^{2}\kappa}$ & & $g_{1}$ & $R_{2}$.4 \\
A.5 & & & & $(0,\infty)$ & 0 & & -1 & 0 &
$\frac{-2+b_{1}}{4 r_{h}^{2}\kappa}$ &
0 & $-\frac{2-b_{1}}{4 r_{h}^{2}\kappa}$ & $g_{1}$ &$R_{3}$.2 \\
A.6 & & & & 1 & $g_{2}$ & & -1 & 0 & $g_{3}$ &
0 & $-g_{3}$ & & $R_{3}$.2\\
A.7 & 0 & & 0 & $(\frac12,\infty)$ & & & -1 & 0 &
$\frac{-2+b_{1}}{4 r_{h}^{2}\kappa}$ &
0 & $\frac{2-b_{1}}{4 r_{h}^{2}\kappa}$ & $g_{1}$ & $R_{3}$.5 \\
\hline
\end{tabular}
\caption{Parameter values that satisfy the class-A SFE and the conservation equation (CE). Blank entries indicate unconstrained parameters.}
\label{AT}
\end{table}

\subsubsection{Summary of the results of the analysis}\label{sral}

Several of the 32 resulting branches are qualitatively equivalent. Retaining only the independent cases reduces the analysis to 18 inequivalent branches. Since a single table listing all parameters would be unwieldy, the results are divided into two tables: Table~\ref{PVbXY} summarizes the $(b_{1}, b_{3})$ parameter space together with the associated $\chi$ and $\psi$ parameters, while Table~\ref{PVAPQ} presents the corresponding functions $A_{1}$, $A_{2}$, and the matter-sector quantities $P$, $\rho$, $E$, along with the conservation equation.

\begin{table}[!h]
\centering
\renewcommand{\arraystretch}{1.15}
\begin{tabular}{|c|c|c|c|c|c|c|c|c|}
\hline
\multicolumn{3}{|c|}{NGR} &
\multicolumn{3}{c|}{$\chi$} &
\multicolumn{3}{c|}{$\psi$} \\
\hline
\# & $b_{\on}$ & $b_{\tr}$ &
$u$ & $\chi_{0}$ & $\gamma_{\on}$ &
$v$ & $\psi_{0}$ & $\gamma_{\tw}$ \\
\hline
A.1 & & & $0$ & $n\pi$ & $0$ & $0$ & $0$ & $0$ \\ \hline
A.2 & & & $0$ & $n\pi$ & $0$ & $0$ & $0$ & $0$ \\ \hline
A.3 & & & $0$ & $n\pi$ & $0$ & $0$ & $0$ & $0$ \\ \hline
A.4 & & & $0$ & $n\pi$ & $0$ & $0$ & $0$ & $0$ \\ \hline
B.1 & $0$ & & $(0,1]$ & $0$ & & $0$ & & \\ \hline
B.2 & $0$ & & $(0,1]$ & $0$ & & $0$ & & \\ \hline
C.1 & $-\frac{2}{3}$ & $\frac{2}{3}$ & $0$ & $n\pi$ & & $0$ & $0$ & $0$ \\ \hline
C.2 & & $-b_{\on}$ & $0$ & $n\pi$ & & $0$ & $0$ & $0$ \\ \hline
D.1 & & & $0$ & $n\pi$ & $0$ & $(0,1]$ & $0$ & \\ \hline
D.2 & & & $0$ & $n\pi$ & $0$ & $(0,1]$ & $0$ & \\ \hline
E.1 & & $0$ & $(0,1]$ & & $0$ & $0$ & $0$ & $0$ \\ \hline
E.2 & & $0$ & $\frac12$ & & $0$ & $0$ & $0$ & $0$ \\ \hline
F.1 & & & $(0,1]$ & $0$ & $0$ & $(0,1]$ & & $0$ \\ \hline
F.2 & & & $(0,1]$ & $0$ & $0$ & $\frac12$ & & $0$ \\ \hline
G.1 & & & $(0,1]$ & $0$ & & $0$ & $0$ & $0$ \\ \hline
G.2 & & & $(0,1]$ & $0$ & & $0$ & $0$ & $0$ \\ \hline
H.1 & & & $0$ & $n\pi$ & $0$ & $0$ & $0$ & \\ \hline
H.2 & & & $0$ & $n\pi$ & $0$ & $0$ & $0$ & \\ \hline
\end{tabular}
\caption{Parameter values satisfying the SFE for $\chi$ and $\psi$. Blank entries indicate unconstrained parameters.}
\label{PVbXY}
\end{table}

We now examine the 18 branches to assess their physical viability using the NGR parameter-space analysis of Section~\ref{NGRps}. Within the perturbative framework, we introduce interpretative criteria, noting that mathematical consistency alone does not ensure physical relevance. Since the geometry is encoded in $A_{1}$, $A_{2}$, $\chi$, and $\psi$, and the perturbative expansion is performed near a LH at $r = r_{\h}$, the ansatz~\eqref{pertAXY} fixes the leading-order geometric structure through the condition~\eqref{localH}. The SFE then determine which parameter combinations contribute at leading order, appear only at subleading order, or must vanish.

To identify which branches correspond to physically meaningful geometric configurations, we must examine the critical case $\alpha_{1}=0$. In the ansatz $A_{2}=\epsilon^{-p}/(\alpha_{1}+\alpha_{2}\epsilon)$ with $p>0$, the case $\alpha_{1}=0$ effectively shifts the leading behaviour of $A_{2}$ from $\epsilon^{-p}$ to $\epsilon^{-(p+1)}$. Attempting to absorb this shift by redefining the ansatz with $p'=p+1$ and using the subleading coefficient $\alpha_{2}$ does not resolve the issue: the SFE then force $\alpha_{2}=0$ as well, causing the ansatz to diverge. Physically, this means that the LH ceases to sit at finite $r=r_{\h}$ at leading order — the horizon scales differently in $\epsilon$ and effectively escapes the perturbative description. Branches with $\alpha_{1}=0$ are therefore regarded as unphysical.

An analogous analysis can be performed for $\chi$ and $\psi$ in the cases where $u>0$ or $v>0$. For branches B.1, B.2, G.1, and G.2, where $\chi_{\n}=0$, one might attempt to redefine the ansatz for $\chi$ as $\chi = \gamma_{1}\epsilon^{u'}$ with $u'=u+1$ and restart the analysis. However, the AFE then force $\gamma_{1}=0$, which implies $\chi=0$ for all four branches, restoring the freedom in the parameter $q$ and rendering branch G.2 identical to A.2. Consequently, G.2 is discarded. Branches B.1, B.2, and G.1 survive this step, although they will be evaluated by the other criteria below. An analogous procedure applies to branches D.1 and D.2, where $\psi_{\n}=0$: redefining $\psi = \gamma_{2}\epsilon^{v'}$ with $v'=v+1$ leads to $\gamma_{2}=0$, implying $\psi=0$, and rendering D.2 identical to A.2. Consequently, D.2 is also discarded.

\begin{table}[!h]
\centering
\rotatebox{90}{
\begin{tabular}{|c|c|c|c|c|c|c|c|c|c|c|c|c|c|}
\hline
\multicolumn{1}{|c|}{NGR} &
\multicolumn{3}{c|}{$A_{1}$} &
\multicolumn{3}{c|}{$A_{2}$} &
\multicolumn{3}{c|}{$P$} &
\multicolumn{2}{c|}{$\rho$} &
\multicolumn{1}{c|}{$E$} &
\multicolumn{1}{c|}{CE} \\
\hline
\# & $q$ & $\beta_{1}$ & $\beta_{2}$ & $p$ & $\alpha_{1}$ & $\alpha_{2}$ &
$y$ & $P_{0}$ & $P_{1}$ & $\rho_{0}$ & $\rho_{1}$ & $Q_{0}$ & \# \\
\hline
A.1 & $ $ & $ $ & $ $ & $(0,\infty)$ & $0$ & $ $ &
$0$ & $\frac{-2+b_{1}}{4 r_{h}^{2}\kappa}$ & $ $ &
$-\frac{-2+b_{1}}{4 r_{h}^{2}\kappa}$ & $-\frac{P_{1}(1+q)}{q}$ &
$g_{1}$ & $R_{2}$.1 \\
\hline
A.2 & $ $ & $ $ & $ $ & $1$ & $g_{2}$ & $ $ &
$0$ & $g_{3}$ & $ $ &
$-g_{3}$ & $-\frac{P_{1}(1+q)}{q}$ & $ $ & $R_2$.1 \\
\hline
A.3 & $0$ & $ $ & $g_{4}$ & $\frac12$ & $ $ & $ $ &
$0$ & $g_{5}$ & $g_{6}$ & $g_{7}$ & $ $ & $ $ & $R_2$.3 \\
\hline
A.4 & $0$ & $ $ & $0$ & $(\frac12,\infty)$ & $ $ & $ $ &
$0$ & $\frac{-2+b_{1}}{4 r_{h}^{2}\kappa}$ & $0$ &
$-\frac{-2+b_{1}}{4 r_{h}^{2}\kappa}$ & $ $ &
$g_{1}$ & $R_2$.4 \\
\hline
B.1 & $-p-u$ & $ $ & $ $ & $(0,\infty)$ & $0$ & $ $ &
$0$ & $-\frac{1}{2r_{h}^{2}\kappa}$ & $ $ &
$\frac{1}{2 r_{h}^{2}\kappa}$ &
$-\frac{P_{1}(1-p-u)}{-p-u}$ & $g_{1}$ & $R_2$.1 \\
\hline
B.2 & $-1-u$ & $ $ & $ $ & $1$ & $g_{2}$ & $ $ &
$0$ & $g_{3}$ & $ $ &
$-g_{3}$ & $\frac{P_{1}u}{-1-u}$ & $ $ & $R_2$.1 \\
\hline
C.1 & $0$ & $-\frac12 \beta_{2} r_{h}$ & $ $ & $(0,\frac12)$ & $ $ & $ $ &
$0$ & $-\frac{2}{3 r_{h}^{2}\kappa}$ & $0$ &
$\frac{2}{3 r_{h}^{2}\kappa}$ & $0$ & $g_{1}$ & $R_2$.3 \\
\hline
C.2 & $0$ & $ $ & $0$ & $(0,\frac12)$ & $0$ & $ $ &
$0$ & $\frac{-2+b_{1}}{4 r_{h}^{2}\kappa}$ & $0$ &
$-\frac{-2+b_{1}}{4 r_{h}^{2}\kappa}$ & $0$ &
$g_{1}$ & $R_2$.4 \\
\hline
D.1 & $-p-v$ & $ $ & $ $ & $(0,1]$ & $0$ & $ $ &
$0$ & $\frac{-2+b_{1}}{4 r_{h}^{2}\kappa}$ & $ $ &
$-\frac{-2+b_{1}}{4 r_{h}^{2}\kappa}$ &
$-\frac{P_{1}(1-p-v)}{-p-v}$ & $g_{1}$ & $R_2$.1 \\
\hline
D.2 & $-1-v$ & $ $ & $ $ & $1$ & $g_{2}$ & $ $ &
$0$ & $g_{3}$ & $ $ &
$-g_{3}$ & $\frac{P_{1}v}{-1-v}$ & $ $ & $R_2$.1 \\
\hline
E.1 & $-u$ & $ $ & $-\frac{2 \beta_{1}}{r_{h}}$ & $1$ & $g_{2}$ & $0$ &
$0$ & $g_{3}$ & $ $ &
$-g_{3}$ & $\frac{P_{1}(1-u)}{u}$ & $ $ & $R_2$.1 \\
\hline
E.2 & $0$ & $ $ & $-\frac{2 \beta_{1}}{r_{h}}$ & $\frac12$ & $h_{1}$ & $0$ &
$0$ & $h_{2}$ & $h_{3}$ &
$h_{4}$ & $ $ & $ $ & $R_2$.3 \\
\hline
F.1 & $-v$ & $ $ & $-\frac{2 \beta_{1}}{r_{h}}$ & $1$ & $g_{2}$ & $0$ &
$0$ & $g_{3}$ & $ $ &
$-g_{3}$ & $\frac{P_{1}(1-v)}{v}$ & $ $ & $R_2$.1 \\
\hline
F.2 & $0$ & $ $ & $-\frac{2 \beta_{1}}{r_{h}}$ & $\frac12$ & $h_{5}$ & $0$ &
$0$ & $h_{6}$ & $h_{7}$ &
$h_{8}$ & $ $ & $ $ & $R_2$.3 \\
\hline
G.1 & $-p-u$ & $ $ & $ $ & $(0,1]$ & $0$ & $ $ &
$0$ & $\frac{-2+b_{1}}{4 r_{h}^{2}\kappa}$ & $ $ &
$-\frac{-2+b_{1}}{4 r_{h}^{2}\kappa}$ &
$\frac{P_{1}(1-p-u)}{p+u}$ & $g_{1}$ & $R_2$.1 \\
\hline
G.2 & $-1-u$ & $ $ & $ $ & $1$ & $g_{2}$ & $ $ &
$0$ & $g_{3}$ & $ $ &
$-g_{3}$ & $\frac{P_{1}u}{-1-u}$ & $ $ & $R_2$.1 \\
\hline
H.1 & $-p$ & $ $ & $a \beta_{1}$ & $(0,1]$ & $0$ & $0$ &
$0$ & $\frac{-2+b_{1}}{4 r_{h}^{2}\kappa}$ & $ $ &
$-\frac{-2+b_{1}}{4 r_{h}^{2}\kappa}$ &
$\frac{P_{1}(1-p)}{p}$ & $g_{1}$ & $R_2$.1 \\
\hline
H.2 & $-1$ & $ $ & $a \beta_{1}$ & $1$ & $g_{2}$ & $-g_{2}(a+\frac{2}{r_{h}})$ &
$0$ & $g_{3}$ & $ $ &
$-g_{3}$ & $0$ & $ $ & $R_2$.1 \\
\hline
\end{tabular}}
\caption{Parameter values of $A_{1}$, $A_{2}$, $P$, $\rho$, and $E$ satisfying the SFE and the conservation equation (CE). Here, $a$ is an arbitrary constant. Blank entries indicate unconstrained parameters.}
\label{PVAPQ}
\end{table}

We can now identify the unphysical branches in Tables~\ref{PVbXY} and~\ref{PVAPQ} and discard them based on the information at hand:
\begin{itemize}
\item Branches with $\alpha_{1}=0$ (leading-order geometric inconsistency): A.1, B.1, C.2, D.1, G.1, and H.1. 
\item Branches equivalent to an already-identified branch under an exponent shift: G.2 and D.2, both of which reduce to A.2.
\item Branches lying outside the physically viable NGR parameter space (Section~\ref{NGRps}): C.1 belongs to Type~II, while E.1 and E.2 belong to Type~I when $b_{1}\neq 0$.
\item Branches that effectively reduce to TEGR and therefore introduce no new NGR behaviour: B.2, and E.1--E.2 when $b_{1}=0$.
\end{itemize}
In total, 12 branches are discarded, leaving 6 cases of potential physical relevance: 
namely, A.2, A.3, A.4, F.1, F.2, and H.2. To further assess these remaining cases, we now 
analyze the next-to-leading-order contributions.

\subsection{Analysis of remaining cases}

We begin this subsection with the 6 candidate branches (A.2, A.3, A.4, F.1, F.2, H.2) that survived the leading-order analysis, and examine the next-to-leading-order AFE to further constrain the parameters $\chi$ and $\psi$.

For all these branches, $\chi$ reduces to a constant. In branches A.2, A.3, A.4, and H.2, we find $\chi = n\pi$ with $n\in\mathbb{Z}$, whereas in branches F.1 and F.2 we have $\chi = 0$. Both choices satisfy the AFE $W_{\al\theta\phi\ar}$ for all branches. Likewise, $\psi$ vanishes except in branches F.1, F.2, and H.2. To determine the full set of parameter values, we return to the AFE and, using the ansatz~\eqref{pertAXY} in Eqs.~\eqref{FEAp}, extend the analysis to next-to-leading order in the perturbative parameter $\epsilon$. Substituting the parameter values listed in Tables~\ref{PVbXY} and~\ref{PVAPQ} for branches F.1, F.2, and H.2, we obtain the following expressions for $W_{\al tr\ar}$:

\begin{equation}
\text{F.1} :\,\frac{b_{\on}\epsilon\psi_{\n}}{g_{\tw}r_{\h}^{2}}= 0\, ,
\qquad
\text{F.2} :\,-\frac{3b_{\on}\sqrt{\epsilon}\,\psi_{\n}}{2r_{\h}^{2}}= 0 \, ,
\qquad
\text{H.2} : \,-\frac{b_{\on}\epsilon\gamma_{\tw}\!\left(3+a r_{\h}(2+a r_{\h})\right)}{r_{\h}^{2}}= 0 \, .
\end{equation}

For branches F.1 and F.2, it is evident that the leading-order terms require $\psi_{\n}=0$. This also restores the freedom in the parameter $q$ for branch F.1, making it a particular case of the more general branch A.2, so F.1 is discarded. In the case of H.2, the leading-order equation vanishes if either $b_{1}=0$ (which corresponds to TEGR and is therefore discarded), $\gamma_{\tw}=0$, or the factor $\bigl(3 + a r_{\h}(2 + a r_{\h})\bigr) = 0$. The latter factor has no real roots for $a$, so the only viable option is $\gamma_{\tw}=0$.

Altogether, these results allow us to conclude that, for all five remaining branches, $\chi = \chi_{\n}$ and $\psi = \psi_{\n}$ are constants. Table~\ref{remain} lists these five cases, with the values of $\chi$ and $\psi$ now fully determined. Blank entries indicate the absence of a constraint on the corresponding parameter. We now re-examine the SFE for these cases, ensuring that the next-to-leading-order terms also satisfy the equations. This refined analysis provides additional information that will help us assess the physical relevance of the remaining cases. Recall that these branches were originally obtained by considering only the leading-order terms (i.e., $\mathcal{O}(\epsilon^{w})$ for $w \leq 0$); we now extend the analysis to the range $0 < w \leq 1$, which corresponds to the first subleading contributions.

\begin{table}[!h]
\centering
\rotatebox{90}{
\begin{tabular}{|c|c|c|c|c|c|c|c|c|c|c|c|c|c|c|c|c|}
\hline
\multicolumn{3}{|c|}{NGR} &
\multicolumn{1}{c|}{$\chi$} &
\multicolumn{1}{c|}{$\psi$} &
\multicolumn{3}{c|}{$A_{1}$} &
\multicolumn{3}{c|}{$A_{2}$} &
\multicolumn{3}{c|}{$P$} &
\multicolumn{2}{c|}{$\rho$} &
\multicolumn{1}{c|}{$E$} \\
\hline
\# & $b_{1}$ & $b_{3}$ & $\chi_{0}$ & $\psi_{0}$ & $q$ & $\beta_{1}$ & $\beta_{2}$ &
$p$ & $\alpha_{1}$ & $\alpha_{2}$ & $y$ & $P_{0}$ & $P_{1}$ &
$\rho_{0}$ & $\rho_{1}$ & $Q_{0}$ \\
\hline
A.2 &  &  & $n\pi$ & $0$ &  &  &  & $1$ & $g_{2}$ &  &
$0$ & $g_{3}$ &  & $-g_{3}$ & $-\frac{P_{1}(1+q)}{q}$ &  \\
\hline
A.3 &  &  & $n\pi$ & $0$ & $0$ &  & $g_{4}$ &
$\frac12$ &  &  & $0$ & $g_{5}$ & $g_{6}$ & $g_{7}$ &  &  \\
\hline
A.4 &  &  & $n\pi$ & $0$ & $0$ &  & $0$ &
$(\frac12,\infty)$ &  &  & $0$ &
$\frac{-2+b_{1}}{4 r_{h}^{2}\kappa}$ & $0$ &
$-\frac{-2+b_{1}}{4 r_{h}^{2}\kappa}$ &  & $g_{1}$ \\
\hline
F.2 &  &  & $0$ & $0$ & $0$ &  & $-\frac{2\beta_{1}}{r_{h}}$ &
$\frac12$ & $h_{5}$ & $0$ & $0$ & $h_{6}$ & $h_{7}$ & $h_{8}$ &  &  \\
\hline
H.2 &  &  & $n\pi$ & $0$ & $-1$ &  & $a\beta_{1}$ &
$1$ & $g_{2}$ & $-g_{2}(a+\frac{2}{r_{h}})$ &
$0$ & $g_{3}$ &  & $-g_{3}$ & $0$ &  \\
\hline
\end{tabular}}
\caption{Parameter values for the five remaining branches. Blank entries indicate unconstrained parameters.}
\label{remain}
\end{table}

\subsubsection{Reviewing A.2}

Let us begin by analyzing case A.2. Substituting the parameter values listed in Table~\ref{remain} into the perturbative SFE, we obtain:
\begin{subequations}\label{A2FE}
\begin{dmath}\label{A2wtt}
W^{t}{}_{t} : \, -\frac{b_{\on}\epsilon g_{2}(1+q)(\beta_{\tw}g_{2}+\alpha_{\tw}\beta_{\on}q)}{\beta_{\on}}+\frac{\epsilon g_{2}^{2}\!\left(2+b_{\on}-2b_{\on}q\right)}{r_{\h}}
\linebreak +\frac{\epsilon\!\left(2+b_{\on}\!\left(-1+2(-1)^{n}g_{2}r_{\h}\right)\right)}{r_{\h}^{3}}\mathrel{\nobreak=}\frac{\epsilon P_{1}(1+q)\kappa}{q} \, ,
\end{dmath}
\vskip -12pt
\begin{dmath}\label{A2wrr}
W^{r}{}_{r} : \, -\frac{\epsilon b_{\on}g_{2}q(\beta_{\tw}g_{2}+\alpha_{\tw}\beta_{\on}q)}{\beta_{\on}}+\frac{\epsilon(2-b_{\on})}{r_{\h}^{3}}+\frac{\epsilon(2+b_{\on})g_{2}^{2}q}{r_{\h}} \mathrel{\nobreak=}\epsilon P_{1}\kappa \, ,
\end{dmath}
\vskip -12pt
\begin{dmath}\label{A2wthth}
W^{\theta}{}_{\theta} : \,\frac{\epsilon g_{2}\!\left(2+b_{\on}+4(1+b_{\on})q\right)\!\left(\beta_{\tw}g_{2}+\alpha_{\tw}\beta_{\on}q\right)}{2\beta_{\on}}-\frac{(-2+b_{\on})\epsilon g_{2}^{2}(1+q)}{2r_{\h}}+\frac{(-1)^{n}b_{\on}\epsilon\!\left(-\beta_{\tw}g_{2}r_{\h}+\beta_{\on}q(g_{2}-\alpha_{\tw}r_{\h})\right)}{\beta_{\on}r_{\h}^{2}}\mathrel{\nobreak=}\epsilon P_{1}\kappa \, .
\end{dmath}
\end{subequations}

We also need to take into account the next-to-leading-order contribution in the conservation equation \eqref{conper}, which, after substituting the parameter values for branch A.2, reduces to:
\begin{equation}\label{conA2}
-\frac{\beta_{\tw} \epsilon P_1}{\beta_{\on} q} = 0 \, .
\end{equation}
Since all terms of order $\mathcal{O}(\epsilon^{0})$ are automatically satisfied by the specific values of the functions $g_{i}$, we focus on the next order, $\mathcal{O}(\epsilon^{1})$. We now explore whether these equations can be satisfied using the freedom in the parameters $q$, $\beta_{\on}$, $\beta_{\tw}$, and $\alpha_{\tw}$, taking into account therestrictions $\alpha_{\on} = g_{2} \neq 0$ and $q \neq 0$. The first indication comes from~\eqref{conA2}, which implies either $\beta_{\tw} = 0$ or $P_{1} = 0$. We analyze both options, together with the combined case $\beta_{\tw} = P_{1} = 0$ and the electrovacuum limit $P = \rho = 0$.

\paragraph{1.} Let us begin by examining the branch $\beta_{\tw}=0$. Since the right-hand sides of Eqs.~\eqref{A2wrr} and~\eqref{A2wthth} coincide at this order (both equal $\epsilon P_{1}\kappa$), we subtract the two equations to eliminate $P_{1}$, which determines $\alpha_{\tw}$:
\begin{dmath}
\alpha_{\tw} =\frac{4 + 2 g_{2}^{2}(-1+q)r_{\h}^{2}+ b_{\on}\!\left[-2 + g_{2}r_{\h}\!\left(-2(-1)^{n}q + g_{2}r_{\h} + 3g_{2}qr_{\h}\right)\right]}{q\,r_{\h}^{2}\!\left[-2(-1)^{n}b_{\on} + g_{2}r_{\h}(2+b_{\on}+4q+6b_{\on}q)\right]} \, .
\end{dmath}
With this value for $\alpha_{\tw}$, we can now obtain an explicit expression for $P_{1}$ either from~\eqref{A2wrr} or~\eqref{A2wthth}, and substitute that result into~\eqref{A2wtt} to obtain:
\begin{equation}\label{gu2con}
g_{2} = \frac{(-1)^{n} b_{\on} \pm 2\sqrt{-1 + (-1 + b_{\on})\,b_{\on}}}{(2 + 3 b_{\on})\,q\,r_{\h}} \, .
\end{equation}
From Eq.~\eqref{gu2v} we observe that this condition implies $Q_{0} = 0$: the branch admits no electric charge. Moreover, in order for $g_{2}$ to remain real and finite, the denominator $(2+3b_{\on})$ must be non-zero (i.e., $b_{\on} \neq -2/3$) and the discriminant must be non-negative ($b_{\on}^{2} - b_{\on} - 1 \geq 0$). Combining these conditions yields:
\begin{equation}\label{bA2}
b_{\on} < -\frac{2}{3} 
\quad \text{or} \quad 
-\frac{2}{3} < b_{\on} \le \frac{1}{2}\left(1 - \sqrt{5}\right)
\quad \text{or} \quad 
b_{\on} \ge \frac{1}{2}\left(1 + \sqrt{5}\right)\, .
\end{equation}
This implies that the models correspond either to a Type~I or a Type~III theory, according to Table~\ref{GFC}. To guarantee a ghost-free model, we must impose the condition $b_{\tr}=-b_{\on}$ for the middle case of~\eqref{bA2}, namely for $-2/3 < b_{\on} \lesssim -0.618$. Under this requirement, the limit $b_{\on}\to 0$ cannot be taken, and therefore the TEGR results cannot be recovered. Consequently, this branch cannot be interpreted as a small correction to TEGR but would instead describe a genuinely distinct theory with no Newtonian limit. Under our physical viability criteria (Section~\ref{NGRps}), the branch is therefore discarded.

\paragraph{2.} Let us now consider the other branch, $P_{1}=0$. In this case, the matter contribution at order $\mathcal{O}(\epsilon^{1})$ vanishes in the SFE, and from~\eqref{A2wrr} we obtain
\begin{equation}
\beta_{\tw} = 
\frac{\beta_{\on}\!\left(2 - b_{\on} + g_{2} q r_{\h}^{2}\!\left[(2 + b_{\on}) g_{2} - \alpha_{\tw} b_{\on} q r_{\h}\right]
\right)}{b_{\on} g_{2}^{2} q r_{\h}^{3}} \, .
\end{equation}
Substituting this into~\eqref{A2wthth}, we recover the same condition for $g_{2}$ as in~\eqref{gu2con}, and therefore Eq.~\eqref{bA2}. This forces us to discard this branch due to its incompatibility with the TEGR limit.

\paragraph{3.} We now examine the branch in which both $\beta_{\tw}$ and $P_{1}$ vanish. Solving Eqs.~\eqref{A2wtt} and~\eqref{A2wrr} in this case leads to exactly the same conclusion as in the previously examined branches: the allowed parameter ranges exclude the TEGR limit. We therefore also discard this branch.

\paragraph{4.} Let us consider the purely electrovacuum limit $P=\rho=0$ with $Q_{0}\neq 0$, which in GR corresponds to the Reissner--Nordström geometry. Referring to Table~\ref{remain}, this choice leads to the following conditions:
\begin{equation}
g_{3}=0\, , \qquad P_{1}=0 \, , \quad \text{and} \quad -\frac{P_{1}(1+q)}{q}=0 \, ,
\end{equation}
where the second condition immediately implies the last one. From the expression for $g_{3}$ in~\eqref{gu3v}, we observe that requiring it to vanish imposes the following possible values for $Q_{0}$:
\begin{equation} \label{Qvo}
Q_{0} = 
\pm\frac{2\sqrt{2\pi(2 - b_{\on})(2 + 3 b_{\on})}\,r_{\h}}
{(2 + b_{\on})\,\sqrt{\kappa}}
\quad \text{or} \quad
Q_{0} = 
\pm\frac{2\sqrt{2\pi}\,r_{\h}}{\sqrt{\kappa}}
\end{equation}
From the first expression for $Q_{0}$, requiring the expression under the square root to be non-negative yields the constraint $-2/3 \leq b_{\on} \leq 2$. For these values of $Q_{0}$, the function $g_{2}$ becomes
\begin{equation}\label{gu2A2n}
g_{2}=\frac{(-1)^{n} b_{\on} (2 + b_{\on})
+ 2(-1)^{m}  \left[-2 + (-3 + b_{\on}) b_{\on}\right]}
{(2 + b_{\on})(2 + 3 b_{\on})\,q\,r_{\h}}
\quad \text{or} \quad
g_{2}=\frac{(-1)^{n} b_{\on} + 2(-1)^{m}(1 + b_{\on})}
{(2 + 3 b_{\on})\,q\,r_{\h}}\, ,
\end{equation}
respectively. Here, the integer $n$ arises from the choice $\chi = n\pi$, and we have introduced $m$ in the factor $(-1)^{m}$, rather than writing an explicit $\pm$, in order to keep track of the different branches. In particular, each expression for $g_{2}$ in~\eqref{gu2A2n} contains four distinct branches corresponding to the sign choices in the numerator.

We now examine all these branches in the SFE. For the first expression of $g_{2}$ in~\eqref{gu2A2n}, the analysis naturally separates into two groups of branches, distinguished by the parity of $m - n$. The first group corresponds to $m - n = 2l$ (same parity); the second to $m - n = 2l + 1$ (opposite parity):
\begin{subequations}
\begin{equation}
m-n=2l: \quad b_{\on} = -\frac{2}{3} \quad \text{and} \quad \beta_{\tw} = -\frac{\beta_{\on}\,(4 +(-1)^{l} q(2 + \alpha_{\tw} q r_{\h}^{2}))}{2 r_{\h}}, \quad \text{or} \quad b_{\on}= 2 \quad \text{and} \quad  \alpha_{\tw} = 0,
\end{equation}
\begin{equation}
m-n=2l+1: \quad b_{\on} = 2, \quad \beta_{\tw} = \beta_{\on}\,\left(\frac{2}{r_{\h}} + (-1)^{l}4 \alpha_{\tw} r_{\h}\right), \quad q = 1   .
\end{equation}
\end{subequations}
These results show that no physical model arises in the absence of a fluid: the allowed values $b_{\on} = -2/3$ (Type~II, DNPS-2) and $b_{\on} = 2$ (no Newtonian limit) are both unphysical. For the second expression of $g_{2}$ in~\eqref{gu2A2n}, we find that none of its possible branches yield a solution to the SFE. Taken together, these findings lead us to conclude that branch A.2 lacks physical relevance, and we therefore discard it.

\subsubsection{Reviewing A.3}

We now analyze case A.3. Substituting the parameter values listed in Table~\ref{remain} into the perturbative SFE, we obtain:
\begin{subequations}
\begin{dmath}\label{A3wtt}
W^{t}{}_{t} : \, \frac{2(-1)^n \alpha_{\on} b_{\on} \sqrt{\epsilon}}{r_{\h}^{2}}-\frac{(-2 + b_{\on})\,\epsilon}{r_{\h}^{3}}  + \frac{\alpha_{\on}^{2} b_{\on} \epsilon\,(\beta_{\on} - g_{4} r_{\h})^{2}}{\beta_{\on}^{2} r_{\h}^{2}} + \frac{2\alpha_{\on}\alpha_{\tw}\epsilon\,\big[(2 + b_{\on})\beta_{\on} - b_{\on} g_{4} r_{\h}\big]}{\beta_{\on} r_{\h}} \mathrel{\nobreak=} -\,\epsilon\,\kappa\,\rho_{1} \, ,
\end{dmath}
\vskip -12pt
\begin{dmath}
W^{r}{}_{r} : \, -\frac{(-2 + b_{\on})\,\epsilon\,(2 + \alpha_{\on}^{2} r_{\h})}{2 r_{\h}^{3}} + \frac{\alpha_{\on}^{2}\,\epsilon\,g_{4}\,\big[2(2 + b_{\on})\beta_{\on} - b_{\on} g_{4} r_{\h}\big]}{2 \beta_{\on}^{2} r_{\h}} \mathrel{\nobreak=} \epsilon\,g_{6}\,\kappa\, ,
\end{dmath}
\vskip -12pt
\begin{dmath}\label{A3wthth}
W^{\theta}{}_{\theta} : \,  \frac{(-1)^{n+1} \alpha_{\on} b_{\on} \sqrt{\epsilon}\, g_{4}}{\beta_{\on} r_{\h}}+\frac{\alpha_{\on}\alpha_{\tw}(2 + b_{\on})\,\epsilon\,g_{4}}{\beta_{\on}}  + \frac{\alpha_{\on}(-2 + b_{\on})\,\epsilon\,\big[-4\alpha_{\tw}\beta_{\on}^{2} r_{\h} + \alpha_{\on}(\beta_{\on} - g_{4} r_{\h})^{2}\big]}{4\beta_{\on}^{2} r_{\h}^{2}} = \epsilon\,g_{6}\,\kappa \, .
\end{dmath}
\end{subequations}
Taking into account the next-to-leading-order contribution in the conservation equation~\eqref{conper} and using the parameter values associated with branch A.3, we obtain
\begin{equation}
\frac{\epsilon g_{4} \left[-g_{4}(g_{5} + g_{7}) + \beta_{\on}(g_{6} + \rho_{1})\right]}{\beta_{\on}^2} = 0.
\end{equation}
Since all terms at order $\mathcal{O}(\epsilon^{0})$ are satisfied by the specific values of the functions $g_{i}$, we focus exclusively on the next orders, namely the $\mathcal{O}(\epsilon^{1/2})$ and $\mathcal{O}(\epsilon^{1})$ contributions.

\paragraph{1.} From Eq.~\eqref{A3wtt}, the leading-order contribution (the $\mathcal{O}(\epsilon^{1/2})$ term) forces either $b_{\on}=0$ or $\alpha_{\on}=0$. The former corresponds to the TEGR limit and is therefore discarded; the latter was shown to be unphysical in the leading-order analysis (see Section~\ref{sral}). Since both options are 
excluded on independent physical grounds, the branch is inconsistent at leading order.

\paragraph{2.} Consider the purely electrovacuum limit $P = 0, \rho = 0$ with $Q_{0}\neq 0$ (the Reissner--Nordström analogue introduced in the A.2 analysis). Referring to Table~\ref{remain}, this choice implies
\begin{equation}
g_{5}=0, \quad g_{6}=0, \quad g_{7}=0, \quad \text{and} \quad \rho_{1}=0 \, .
\end{equation}
Examining these conditions using the expressions for $g_{5}$, $g_{6}$, and $g_{7}$ given in~\eqref{gfun} shows that $\alpha_{\on}$ and $Q_{0}$ must satisfy
\begin{equation}
\alpha_{\on}=\pm \sqrt{\frac{(2-b_{\on})b_{\on}}{(2+3b_{\on})r_{\h}}}
\quad \text{and} \quad
Q_{0}=\pm \frac{2 r_{\h}\sqrt{\pi(2-b_{\on})}}{\sqrt{\kappa}} \, .
\end{equation}
Requiring these quantities to be real and finite forces the parameter $b_{\on}$ to satisfy
\begin{equation}\label{bA3}
b_{\on}<-\frac{2}{3}
\quad \text{or} \quad
0 \leq b_{\on} \leq 2 \, .
\end{equation}
These ranges lie outside the physically viable NGR parameter space specified in~\eqref{bphy}, with the sole exception of $b_{\on}=0$, which corresponds to the TEGR case. Therefore, no physical NGR solution exists in the absence of a fluid. Taken together, these results render branch A.3 unphysical, and it is therefore discarded.  

\subsubsection{Reviewing A.4}

We now analyze case A.4. Substituting the parameter values listed in Table~\ref{remain} into the perturbative framework, we find that the conservation equation is identically satisfied (since $q=0$ and $P_{1}=\rho_{1}=0$ remove all non-trivial contributions), and the SFE reduce to
\begin{subequations}
\begin{dmath}\label{A4wtt}
W^{t}{}_{t} :\, -\frac{(-2 + b_{\on})\,\epsilon}{r_{\h}^{3}} + \frac{2(-1)^{n}\alpha_{\on} b_{\on}\,\epsilon^{p}}{r_{\h}^{2}} + \frac{\alpha_{\on}^{2}(2 + b_{\on})\,\epsilon^{-1 + 2p}\,p}{r_{\h}} \mathrel{\nobreak=} -\,\epsilon\,\kappa\,\rho_{1} \, , 
\end{dmath}
\vskip -12pt
\begin{dmath}\label{A4wrr}
W^{r}{}_{r} :\, -\frac{(-2 + b_{\on})\,\epsilon}{r_{\h}^{3}} - \frac{\alpha_{\on}^{2}(-2 + b_{\on})\,\epsilon^{2p}}{2 r_{\h}^{2}} \mathrel{\nobreak=} 0\, , 
\end{dmath}
\vskip -12pt
\begin{dmath}\label{A4wthth}
W^{\theta}{}_{\theta} :\, -\frac{\alpha_{\on}^{2}(-2 + b_{\on})\,\epsilon^{-1 + 2p}p}{2r_{\h}} + \frac{\alpha_{\on}(-2 + b_{\on})\,\epsilon^{2p}\big(\alpha_{\on}p - \alpha_{\tw}(1 + 2p)r_{\h}\big)}{2r_{\h}^{2}} \mathrel{\nobreak=} 0\, .
\end{dmath}
\end{subequations}
The leading terms in the SFE appear at orders $\mathcal{O}(\epsilon^{-1+2p})$, $\mathcal{O}(\epsilon^{p})$, $\mathcal{O}(\epsilon^{2p})$, and $\mathcal{O}(\epsilon^{1})$. Depending on the value of $p$, different terms become leading, so we analyze the cases $1/2 < p < 1$ and $p \geq 1$ separately.

\paragraph{1.} In the regime $1/2 < p < 1$, the leading subleading term in~\eqref{A4wtt} is the $\mathcal{O}(\epsilon^{p})$ term, which is proportional to $\alpha_{\on} b_{\on}$. Requiring it to vanish forces either $\alpha_{\on}=0$ (unphysical) or $b_{\on}=0$ (TEGR, discarded). For $p \geq 1$, the $\mathcal{O}(\epsilon^{2p})$ terms in~\eqref{A4wrr} and~\eqref{A4wthth} are subleading, and the only $\mathcal{O}(\epsilon^{1})$ term in~\eqref{A4wrr} vanishes only if $b_{\on}=2$, which characterizes a theory lacking a Newtonian limit. The branch is therefore discarded in both subregimes.

\paragraph{2.} Consider the purely electrovacuum limit $P=0,\,\rho=0$ (the Reissner--Nordström analogue). Table~\ref{remain} gives $P_{0}=(-2+b_{\on})/(4 r_{\h}^{2}\kappa)$ for branch A.4; setting $P_{0}=0$ forces $b_{\on}=2$. This value is unphysical (no Newtonian limit), and therefore the branch must be discarded.

Taken together, these results show that branch A.4 admits no physically viable configurations, and it is therefore discarded.

\subsubsection{Reviewing F.2}

We now analyze case F.2. Substituting the parameter values listed in Table~\ref{remain} into the perturbative framework, the SFE take the form:
\begin{subequations}\label{SFEF2}
\begin{dmath}\label{F2wtt}
W^{t}{}_{t} : \,  \frac{2 b_{\on} \sqrt{\epsilon}\,h_{5}}{r_{\h}^{2}} + \frac{\epsilon\big(2 + b_{\on}(-1 + 9 h_{5}^{2} r_{\h})\big)}{r_{\h}^{3}} \mathrel{\nobreak=} -\epsilon\,\kappa\,\rho_{1} \, ,
\end{dmath}
\vskip -12pt
\begin{dmath}\label{F2wrr}
W^{r}{}_{r} : \,  -\frac{\epsilon\big(2(-2 + b_{\on}) + 3(2 + 3 b_{\on})h_{5}^{2}r_{\h}\big)}{2r_{\h}^{3}} \mathrel{\nobreak=} \epsilon\,h_{7}\,\kappa \, , 
\end{dmath}
\vskip -12pt
\begin{dmath}\label{F2wthth}
W^{\theta}{}_{\theta}: \, \frac{2 b_{\on} \sqrt{\epsilon}\,h_{5}}{r_{\h}^{2}} + \frac{9(-2 + b_{\on})\,\epsilon\,h_{5}^{2}}{4r_{\h}^{2}} \mathrel{\nobreak=} \epsilon\,h_{7}\,\kappa \, . 
\end{dmath}
\end{subequations}
The conservation equation~\eqref{conper}, rewritten using the parameter values associated with branch F.2, yields
\begin{equation}\label{conF2}
-\frac{2\epsilon\big(2h_{6} + 2h_{8} + r_{\h}(h_{7} + \rho_{1})\big)}{r_{\h}^{2}} = 0 \, . 
\end{equation}
Since all terms at order $\mathcal{O}(\epsilon^{0})$ in the system~\eqref{SFEF2} and~\eqref{conF2} are satisfied, we now focus on the next order; namely, the $\mathcal{O}(\epsilon^{1/2})$ contributions, which as we will see are sufficient to close the case.

\paragraph{1.} The $\mathcal{O}(\epsilon^{1/2})$ contribution in Eqs.~\eqref{F2wtt} and~\eqref{F2wthth} is proportional to $b_{\on} h_{5}$. Requiring it to vanish forces either $b_{\on}=0$ (TEGR, discarded) or $h_{5}=0$. Substituting the F.2 parameter values into the definition of $h_{5}$ given in~\eqref{hu5v}, the condition $h_{5}=0$ implies:
\begin{equation}
b_{\on}=2 \quad \text{and} \quad Q_{0}=0, \quad \text{or} \quad Q_{0} = 
\pm\frac{r_{\h}\sqrt{2\pi(2-b_{\on})}}{\sqrt{\kappa}}\, . 
\end{equation}
The first set of values is discarded since $b_{\on}=2$ rules out the Newtonian limit. The second set makes not only $h_{5}=0$ but also $h_{7}=0$, as is evident from~\eqref{hu7v}. With $h_{5}=h_{7}=0$, Eq.~\eqref{F2wthth} is automatically satisfied, but the only surviving term in Eq.~\eqref{F2wrr} forces $b_{\on}=2$, which again rules out the Newtonian limit. We therefore discard this branch as well.

\paragraph{2.} Consider the purely electrovacuum limit $P=0,\,\rho=0$ (the Reissner--Nordström analogue). Referring to Table~\ref{remain}, this choice implies
\begin{equation}
h_{6}=0, \quad h_{7}=0, \quad h_{8}=0, \quad \text{and} \quad \rho_{1}=0 \, .
\end{equation}
Examining these conditions using the expressions for $h_{6}$, $h_{7}$, and $h_{8}$ given in~\eqref{hfun} shows that $b_{\on}=2$ and $Q_{0}=0$, thereby rendering this branch unphysical.

Taken together, these results render branch F.2 unphysical, and it is therefore discarded.
 
\subsubsection{Reviewing H.2}

We now analyze the final case, H.2. Substituting the parameter values listed in Table~\ref{remain} into the perturbative framework, the SFE take the form:
\begin{subequations}
\begin{dmath}\label{H2wtt}
W^{t}{}_{t} : \, \frac{\epsilon\!\left(2 + 2g_{2}^{2}r_{\h}^{2} + b_{\on}\big(-1 + g_{2}r_{\h}(2(-1)^{n} + 3g_{2}r_{\h})\big)\right)}{r_{\h}^{3}} \mathrel{\nobreak=} 0 \, , 
\end{dmath}
\vskip -12pt
\begin{dmath}\label{H2wrr}
W^{r}{}_{r} : \, \frac{\epsilon\!\left(2 - b_{\on} + (-2 + b_{\on})g_{2}^{2}r_{\h}^{2} + 2a b_{\on}g_{2}^{2}r_{\h}^{3}\right)}{r_{\h}^{3}} \mathrel{\nobreak=} \epsilon P_{1}\kappa \, , 
\end{dmath}
\vskip -12pt
\begin{dmath}\label{H2wthth}
W^{\theta}{}_{\theta} : \, \frac{\epsilon\,g_{2}\!\left(- (2 + 3b_{\on})g_{2}r_{\h}(1 + a r_{\h}) - (-1)^{n} b_{\on}(3 + 2a r_{\h})\right)}{r_{\h}^{2}} \mathrel{\nobreak=} \epsilon P_{1}\kappa \, .
\end{dmath}
\end{subequations}
The conservation equation~\eqref{conper}, rewritten using the parameter values associated with branch H.2, yields
\begin{equation}
P_{1}=a \, g_{3} .
\end{equation}
Since all terms at order $\mathcal{O}(\epsilon^{0})$ are satisfied by the specific values of the $g_{i}$ functions, we now focus on the next order, namely the terms of order $\mathcal{O}(\epsilon^{1})$. 

\paragraph{1.} Setting $q = -1$ in Eq.~\eqref{gu2con} (derived in the A.2 analysis but formally identical here) gives the value of $g_{2}$ required by~\eqref{H2wtt}. However, from~\eqref{H2wrr} and~\eqref{H2wthth} we find that $g_{2}$ must instead be given by
\begin{equation}
g_{2}=\frac{-(-1)^{n} b_{\on}(3 + 2a r_{\h}) \pm \sqrt{-16a r_{\h} - 32 b_{\on}(1 + a r_{\h}) + b_{\on}^{2}\!\left(25 + 4a r_{\h}(8 + a r_{\h})\right)}}{2 r_{\h}\!\left(4 b_{\on} + a(2 + 5 b_{\on}) r_{\h}\right)}\, .
\end{equation}
Since there is no value of $a$ for which the two expressions for $g_{2}$ can be reconciled, the system is inconsistent. This inconsistency is stronger than a mere restriction on $a$: the two expressions encode geometrically distinct constraints on the tetrad structure that admit no common solution. No fluid-supported H.2 branch therefore exists at this order.

\paragraph{2.} Consider the purely electrovacuum limit $P=0,\,\rho=0$ (the Reissner--Nordström analogue). Referring to Table~\ref{remain}, this choice implies
\begin{equation}
g_{3}=0 \quad \text{and} \quad P_{1}=0 \, ,
\end{equation}
the latter being automatically implied by the conservation equation $P_{1} = a\, g_{3}$. From the roots of $g_{3}$ we obtain the possible values of $Q_{0}$ given in~\eqref{Qvo}. However, the $\mathcal{O}(\epsilon^{0})$ content of Eq.~\eqref{H2wtt} (absorbed into the specific values of the $g_{i}$ functions) requires $Q_{0} = 0$ for consistency with the electrovacuum limit, which in turn forces either $b_{\on}=2$ (no Newtonian limit) or $b_{\on}=-2/3$ (Type~II, DNPS-2). Both values are unphysical, and the branch must therefore be discarded.

Taken together, these results render branch H.2 unphysical, and it is therefore discarded.

\section{Discussion}

In this paper, we have reviewed the essential features of teleparallel geometry required to construct a well-defined teleparallel theory in a fully covariant framework. Motivated by the freedom in its parameter space, we have examined NGR as a potential deformation or extension of TEGR. Our analysis incorporates several key aspects of well-tested physics, including the Newtonian limit, together with existing results on ghost stability and gravitational-wave propagation~\cite{bahamonde2025revisiting, golovnev2024gravitational}. We have also revisited previous findings on SSS vacuum black hole solutions~\cite{lopez2025black} and extended the analysis to perfect fluid and electrovacuum configurations. The inclusion of matter sources was motivated by the expectation that non-trivial energy--momentum content might relax the algebraic constraints on the NGR parameters imposed by the vacuum SFE, thereby opening a window for physically viable black hole solutions. Altogether, these investigations provide a comprehensive assessment of the physical viability of NGR black holes.

A summary of the cases that satisfy the non-vacuum SSS AFE and SFE at leading perturbative order is given in Tables~\ref{PVbXY} and~\ref{PVAPQ}. With the exception of those branches that reduce trivially to TEGR, all remaining cases exhibit nontrivial constraints on the functions $\chi$ and $\psi$, most commonly fixing $\chi = n\pi$ and $\psi = 0$. The recurrence of this pattern across the non-TEGR branches reflects the strong geometric rigidity imposed by the combined AFE and LH conditions.

Starting from 18 inequivalent branches satisfying the leading-order equations, physical and analytical criteria reduce the pool to 5 candidates to be examined at higher orders. In several instances, the surviving constraints force specific values of $b_{1}$ that lie outside the physically admissible parameter space of NGR. Consequently, although certain branches admit formal solutions at leading order, they become inconsistent or unphysical once higher-order contributions are taken into account.

All cases for which there may exist black hole solutions that do not reduce to TEGR share a number of unfavorable physical features. In particular, they fail one or more of the following viability criteria: (i) the theory must be ghost-free; (ii) the theory must admit a Newtonian limit, and therefore be consistent with solar-system tests; (iii) the theory must support a propagating spin-2 mode, which enables gravitational-wave propagation~\cite{bahamonde2025revisiting, golovnev2024gravitational}.

Altogether, our analysis shows that none of the non-TEGR branches of NGR examined here yield a physically consistent black hole solution, either in vacuum, perfect fluid, or electrovacuum. In all potentially viable cases, the underlying NGR model fails to satisfy one or more essential physical requirements (ghost-freedom, Newtonian limit, or propagating spin-2 modes), and therefore cannot be interpreted as a consistent modified gravity theory admitting black hole solutions. Within the perturbative framework considered, NGR does not admit physically acceptable black-hole solutions distinct from those in TEGR.

This negative result, together with the earlier vacuum analysis of~\cite{lopez2025black} and the analogous findings for $F(T)$ gravity~\cite{coley2025black}, indicates that torsion-based modifications of TEGR within the class examined here cannot generate novel black hole geometries without sacrificing essential physical properties. It is important to emphasize that this result applies to the specific class of NGR models considered here: static, spherically symmetric configurations with a minimally coupled perfect fluid and electromagnetic source, analyzed within a perturbative framework around a local horizon. Whether NGR is \emph{fundamentally} incompatible with astrophysical black holes, or only with this class of configurations, remains an open question. Relaxing any of these assumptions, for example, by considering axially symmetric geometries, time-dependent configurations, non-minimal matter couplings, or higher-derivative extensions of NGR, may yet reveal distinct black-hole-like solutions. Likewise, whether the conclusion extends to the broader family of torsion theories incorporating additional scalar invariants (such as $f(\mathscr{V}, \mathscr{A}, \mathscr{T})$ theories) remains to be explored.

\section*{Acknowledgement}

D. F. L. and B. Y. acknowledges support from the Department of Mathematics and Statistics at Dalhousie University, Canada. A. A. C. is supported by the Natural Sciences and Engineering Research Council of Canada (NSERC).

%
%

\printbibliography

%
%

\appendix

\begin{center}
\vskip 0.7cm
{\Large \bf Appendix \\ }
\vskip 0.7cm
\end{center}

\renewcommand{\theequation}{\Alph{section}.\arabic{equation}}
\setcounter{equation}{0}

\section{Auxiliary functions}\label{AuxF}
The functions \( F_{i} \) introduced in \eqref{FEV}, each depending only on \( A_{1} \) and \( A_{2} \), are listed below:
\begin{subequations}\label{FV}
\begin{dmath} 
F_{1}  = b_{1}[\ln A_{1}]'' + \frac{2+b_{1}}{r^{2}} + b_{1}[\ln A_{1}]'[\ln(A_{1}^{1/2}r^{2}/A_{2})]' + \frac{2+b_{1}}{r}[\ln A_{2}]'- \left(3-5b_{1}/2\right)\frac{1}{r^{2}} + \left(1-b_{1}/2\right)\left(A_{2}/r\right)^{2}, 
\end{dmath}

\begin{dmath}
F_{2} = -\frac{b_{1}}{2}\big([\ln A_{1}]'\big)^{2} + \frac{2+b_{1}}{r}[\ln A_{1}]' + \frac{1-b_{1}/2}{r^{2}}\left(1-A_{2}^{2}\right), 
\end{dmath}

\begin{dmath}
F_{3} = \left(1+b_{1}/2\right)[\ln A_{1}]'' + \frac{1-b_{1}/2}{r^{2}}\left(1+r[\ln(A_{1}r/A_{2})]'\right)+ (1+b_{1})([\ln A_{1}]')^{2}-\left(1+b_{1}/2\right)[\ln A_{1}]'[\ln A_{2}]'. 
\end{dmath}
\end{subequations}
The functions \( G_{i}(\epsilon) \) introduced in \eqref{FEAp} and \eqref{FESp}, expressed in terms of the perturbative parameter \( \epsilon \), are given by:
\begin{subequations}\label{GV}
\begin{dmath}\label{G1}
G_{1}(\epsilon)=\frac{b_{\on}\epsilon^{-1+v}}{2\alpha_{\on}\beta_{\on}r_{\h}}
\!\left[\alpha_{\tw}\beta_{\on}\psi_{\n}r_{\h}v
+\alpha_{\on}\!\left(2\beta_{\on}\psi_{\n}v
+\beta_{\tw}\psi_{\n}r_{\h}v
+\beta_{\on}\gamma_{\tw}r_{\h}(1+v)(p+q+v)\right)\right]
+\frac{b_{\on}\epsilon^{v}}{2\alpha_{\on}^{2}\beta_{\on}^{2}r_{\h}^{2}}
\!\left[-\alpha_{\tw}^{2}\beta_{\on}^{2}\psi_{\n}r_{\h}^{2}v
+\alpha_{\on}\alpha_{\tw}\beta_{\on}^{2}\gamma_{\tw}r_{\h}^{2}(1+v)
+\alpha_{\on}^{2}\!\left(-\beta_{\tw}^{2}\psi_{\n}r_{\h}^{2}v
+\beta_{\on}\beta_{\tw}\gamma_{\tw}r_{\h}^{2}(1+v)
+2\beta_{\on}^{2}(-\psi_{\n}v+\gamma_{\tw}r_{\h}(1+v))\right)\right]
+\frac{b_{\on}\epsilon^{-p}}{\alpha_{\on}^{2}r_{\h}^{3}}
\!\left[-\alpha_{\tw}\epsilon r_{\h}
+\alpha_{\on}(-2\epsilon+r_{\h})\right]
\cos[\epsilon^{u}(\chi_{\n}+\epsilon\gamma_{\on})]
\sinh[\epsilon^{v}(\epsilon\gamma_{\tw}+\psi_{\n})]
+\frac{b_{\tr}\epsilon^{-1-p+u}}{\alpha_{\on}^{2}r_{\h}^{2}}
\!\left[\alpha_{\on}\epsilon\gamma_{\on}(\epsilon-r_{\h})
+\alpha_{\on}(\chi_{\n}\epsilon-(\chi_{\n}+\epsilon\gamma_{\on})r_{\h})u
+\alpha_{\tw}\epsilon r_{\h}(\epsilon\gamma_{\on}+\chi_{\n}u)\right]
\sin[\epsilon^{u}(\chi_{\n}+\epsilon\gamma_{\on})]
\sinh[\epsilon^{v}(\epsilon\gamma_{\tw}+\psi_{\n})] \, ,
\end{dmath}

\begin{dmath}\label{G2}
G_{2}(\epsilon)=\frac{(b_{\on}+b_{\tr})\,\epsilon^{-1+u}}{2\alpha_{\on}\beta_{\on}r_{\h}}
\!\left[\alpha_{\tw}\beta_{\on}\chi_{\n}r_{\h}u
+\alpha_{\on}\!\left(2\beta_{\on}\chi_{\n}u
+\beta_{\tw}\chi_{\n}r_{\h}u
+\beta_{\on}\gamma_{\on}r_{\h}(1+u)(p+q+u)\right)\right]
+\frac{(b_{\on}+b_{\tr})\,\epsilon^{u}}{2\alpha_{\on}^{2}\beta_{\on}^{2}r_{\h}^{2}}
\!\left[-\alpha_{\tw}^{2}\beta_{\on}^{2}\chi_{\n}r_{\h}^{2}u
+\alpha_{\on}\alpha_{\tw}\beta_{\on}^{2}\gamma_{\on}r_{\h}^{2}(1+u)
+\alpha_{\on}^{2}\!\left(-\beta_{\tw}^{2}\chi_{\n}r_{\h}^{2}u
+\beta_{\on}\beta_{\tw}\gamma_{\on}r_{\h}^{2}(1+u)
+2\beta_{\on}^{2}(-\chi_{\n}u+\gamma_{\on}r_{\h}(1+u))\right)\right]
+\frac{\epsilon^{-1-p}}{\alpha_{\on}^{2}\beta_{\on}^{2}r_{\h}^{3}}
\!\left[
2\alpha_{\on}(-b_{\on}+b_{\tr})\beta_{\on}^{2}\epsilon^{2}
+\beta_{\on}\epsilon\!\left(\alpha_{\tw}(-b_{\on}+b_{\tr})\beta_{\on}\epsilon
+\alpha_{\on}(b_{\on}\beta_{\on}+b_{\tr}\beta_{\tw}\epsilon+b_{\tr}\beta_{\on}(-1+q))\right)r_{\h}\right.\\[4pt]
\left.
+b_{\tr}\!\left(\beta_{\tw}\epsilon(-\alpha_{\on}\beta_{\on}
+\alpha_{\tw}\beta_{\on}\epsilon+\alpha_{\on}\beta_{\tw}\epsilon)
+\beta_{\on}^{2}(-\alpha_{\on}+\alpha_{\tw}\epsilon)q\right)r_{\h}^{2}
\right]
\cosh[\epsilon^{v}(\epsilon\gamma_{\tw}+\psi_{\n})]
\sin[\epsilon^{u}(\chi_{\n}+\epsilon\gamma_{\on})]
+\frac{b_{\tr}\epsilon^{-2p}}{\alpha_{\on}^{3}r_{\h}^{3}}
(2\alpha_{\on}\epsilon-\alpha_{\on}r_{\h}+2\alpha_{\tw}\epsilon r_{\h})
\sin[2\epsilon^{u}(\chi_{\n}+\epsilon\gamma_{\on})]
+\frac{b_{\tr}\epsilon^{-1-p+v}}{\alpha_{\on}^{2}r_{\h}^{2}}
\!\left[\alpha_{\on}\epsilon\gamma_{\tw}(\epsilon-r_{\h})
+\alpha_{\on}(\epsilon\psi_{\n}-(\epsilon\gamma_{\tw}+\psi_{\n})r_{\h})v
+\alpha_{\tw}\epsilon r_{\h}(\epsilon\gamma_{\tw}+\psi_{\n}v)\right]
\sin[\epsilon^{u}(\chi_{\n}+\epsilon\gamma_{\on})]
\sinh[\epsilon^{v}(\epsilon\gamma_{\tw}+\psi_{\n})] \, ,
\end{dmath}

\begin{dmath}\label{G3}
G_{3}(\epsilon)=-\frac{1}{2}(b_{\on}+b_{\tr})\,\epsilon^{-2+2u}
\!\left(\epsilon\gamma_{\on}+\chi_{\n}u\right)
\!\left(\chi_{\n}u+\epsilon(\gamma_{\on}+2\gamma_{\on}u)\right)
+\frac{1}{2}b_{\on}\,\epsilon^{-2+2v}
\!\left(\epsilon\gamma_{\tw}+\psi_{\n}v\right)
\!\left(\psi_{\n}v+\epsilon(\gamma_{\tw}+2\gamma_{\tw}v)\right)
-\frac{\epsilon^{-2p}}{2\alpha_{\on}^{2}r_{\h}^{2}}
\!\left(-2+b_{\on}+2b_{\tr}-2b_{\tr}\cos[2\epsilon^{u}(\chi_{\n}+\epsilon\gamma_{\on})]\right)
-\frac{2b_{\on}\epsilon^{-p}}{\alpha_{\on}r_{\h}^{2}}
\cos[\epsilon^{u}(\chi_{\n}+\epsilon\gamma_{\on})]
\cosh[\epsilon^{v}(\epsilon\gamma_{\tw}+\psi_{\n})]
+\frac{2b_{\tr}\epsilon^{-1-p+u}}{\alpha_{\on}^{2}r_{\h}^{2}}
\!\left[-\alpha_{\tw}\chi_{\n}\epsilon r_{\h}u
+\alpha_{\on}\!\left(-\chi_{\n}\epsilon u+\chi_{\n}r_{\h}u+\epsilon\gamma_{\on}r_{\h}(1+u)\right)\right]
\cosh[\epsilon^{v}(\epsilon\gamma_{\tw}+\psi_{\n})]
\sin[\epsilon^{u}(\chi_{\n}+\epsilon\gamma_{\on})] \, ,
\end{dmath}

\begin{dmath}\label{G4}
G_{4}(\epsilon)=-\frac{1}{2}(b_{\on}+b_{\tr})\,\epsilon^{-2+2u}
\!\left(\epsilon\gamma_{\on}+\chi_{\n}u\right)
\!\left(\chi_{\n}u+\epsilon(\gamma_{\on}+2\gamma_{\on}u)\right)
+\frac{1}{2}b_{\on}\,\epsilon^{-2+2v}
\!\left(\epsilon\gamma_{\tw}+\psi_{\n}v\right)
\!\left(\psi_{\n}v+\epsilon(\gamma_{\tw}+2\gamma_{\tw}v)\right)
+\frac{\epsilon^{-2p}}{2\alpha_{\on}^{2}r_{\h}^{2}}
\!\left(-2+b_{\on}+2b_{\tr}-2b_{\tr}
\cos[2\epsilon^{u}(\chi_{\n}+\epsilon\gamma_{\on})]\right) \, ,
\end{dmath}

\begin{dmath}\label{G5}
G_{5}(\epsilon)=\frac{1}{2}(b_{\on}+b_{\tr})\,\epsilon^{-2+2u}
\!\left(\epsilon\gamma_{\on}+\chi_{\n}u\right)
\!\left(\chi_{\n}u+\epsilon(\gamma_{\on}+2\gamma_{\on}u)\right)
-\frac{1}{2}b_{\on}\,\epsilon^{-2+2v}
\!\left(\epsilon\gamma_{\tw}+\psi_{\n}v\right)
\!\left(\psi_{\n}v+\epsilon(\gamma_{\tw}+2\gamma_{\tw}v)\right)
+\frac{b_{\on}\epsilon^{-1-p}}{\alpha_{\on}^{2}\beta_{\on}r_{\h}^{2}}
\!\left[\alpha_{\on}\beta_{\on}\epsilon q
+\alpha_{\tw}\beta_{\on}\epsilon q r_{\h}
-\alpha_{\on}(\beta_{\tw}\epsilon+\beta_{\on}q)r_{\h}\right]
\cos[\epsilon^{u}(\chi_{\n}+\epsilon\gamma_{\on})]
\cosh[\epsilon^{v}(\epsilon\gamma_{\tw}+\psi_{\n})]
+\frac{(b_{\on}-b_{\tr})\epsilon^{-1-p+u}}{\alpha_{\on}^{2}r_{\h}^{2}}
\!\left[-\alpha_{\tw}\chi_{\n}\epsilon r_{\h}u
+\alpha_{\on}\!\left(-\chi_{\n}\epsilon u+\chi_{\n}r_{\h}u+\epsilon\gamma_{\on}r_{\h}(1+u)\right)\right]
\cosh[\epsilon^{v}(\epsilon\gamma_{\tw}+\psi_{\n})]
\sin[\epsilon^{u}(\chi_{\n}+\epsilon\gamma_{\on})]
-\frac{b_{\on}\epsilon^{-1-p+v}}{\alpha_{\on}^{2}r_{\h}^{2}}
\!\left[-\alpha_{\tw}\epsilon\psi_{\n}r_{\h}v
+\alpha_{\on}\!\left(-\epsilon\psi_{\n}v+\psi_{\n}r_{\h}v+\epsilon\gamma_{\tw}r_{\h}(1+v)\right)\right]
\cos[\epsilon^{u}(\chi_{\n}+\epsilon\gamma_{\on})]
\sinh[\epsilon^{v}(\epsilon\gamma_{\tw}+\psi_{\n})] \, .
\end{dmath}
\end{subequations}

\section{Expression catalogue}

This is the list of functions appearing in the parameter values for the matter and geometrical sectors that solve the SFE and the conservation equation when expanded perturbatively up to first order in the parameter $\epsilon$.

\begin{subequations} \label{gfun}
\begin{flalign}
g_{\on} &= \pm r_{\h} \sqrt{(2 - b_{\on})\frac{2\pi}{\kappa}}\, , &
\end{flalign}

\begin{flalign}\label{gu2v}
g_{\tw} &= \dfrac{2(-1)^n b_{\on} \pi r_{\h} \pm \sqrt{2\pi} \sqrt{8(-1 + (-1 + b_{\on})b_{\on})\pi r_{\h}^2 + (2 + 3b_{\on})Q_0^2 \kappa}}{2(2 + 3b_{\on})\pi q\, r_{\h}^2} \, ,&
\end{flalign}

\begin{flalign}\label{gu3v}\nonumber
g_{\tr} &= \dfrac{-8\!\left(4 + b_{\on}(8 + b_{\on} - 2b_{\on}^2)\right)\pi r_{\h}^2
+ (2 + b_{\on})(2 + 3b_{\on}) Q_0^2 \kappa}{8(2 + 3b_{\on})^2 \pi r_{\h}^4 \kappa} \\
& \quad \pm\;\dfrac{4(-1)^{n} b_{\on}^2 \sqrt{2\pi}\, r_{\h}\,
\sqrt{\,8\!\left(-1 + (-1 + b_{\on})b_{\on}\right)\!\pi r_{\h}^2 + (2 + 3b_{\on}) Q_0^2 \kappa}}{8(2 + 3b_{\on})^2 \pi r_{\h}^4 \kappa}\, , &
\end{flalign}

\begin{flalign}
g_{\text{\tiny 4}}& = \dfrac{\beta_{\on} \left[(-2 + b_{\on})\pi r_{\h}^2 \left(2 + \alpha_{\on}^2 r_{\h}\right) + Q_0^2 \kappa \right]}{\alpha_{\on}^2 (2 + b_{\on}) \pi r_{\h}^4}  \, , &
\end{flalign}

\begin{flalign}
g_{\text{\tiny 5}}& = \dfrac{4(-2 + b_{\on})\pi r_{\h}^2 + Q_0^2 \kappa}{8\pi r_{\h}^4 \kappa} \, , &
\end{flalign}

\begin{flalign}
g_{\text{\tiny 6}} &=\dfrac{\left[(-2 + b_{\on})\pi r_{\h}^2 \left(2 + \alpha_{\on}^2 r_{\h}\right) + Q_0^2 \kappa\right]\left[2\pi r_{\h}^2 \left( -(-2 + b_{\on})b_{\on} + \alpha_{\on}^2 (2 + 3b_{\on}) r_{\h} \right) - b_{\on} Q_0^2 \kappa\right]}{2\alpha_{\on}^2 (2 + b_{\on})^2 \pi^2 r_{\h}^8 \kappa} \, , &
\end{flalign}

\begin{flalign}
g_{\text{\tiny 7}} &= 
\dfrac{4\pi r_{\h}^2 \left[(-2 + b_{\on})^2 - 2\alpha_{\on}^2 (2 + 3b_{\on}) r_{\h} \right] 
+ (-2 + 3b_{\on}) Q_0^2 \kappa}{8(2 + b_{\on})\pi r_{\h}^4 \kappa} \, , &
\end{flalign}
\end{subequations}
\begin{subequations}\label{hfun}
\begin{flalign}
h_{\on} &=\pm \frac{ \sqrt{2(-2 + b_{\on})\pi r_{\h}^2 + Q_0^2 \kappa}}{ \sqrt{\pi r_{\h}^3 \left(-2 + b_{\on}(-3 + \chi_{\n}^2 r_{\h})\right)}} \, , &
\end{flalign}

\begin{flalign}
h_{\tw} &=\frac{2(-2 + b_{\on})\pi r_{\h}^2 \left(-4 + b_{\on}(-6 + \chi_{\n}^2 r_{\h})\right) 
- (2 + 3b_{\on}) Q_0^2 \kappa}{8\pi r_{\h}^4 \left(-2 + b_{\on}(-3 + \chi_{\n}^2 r_{\h})\right)\kappa}\, , &
\end{flalign}

\begin{flalign}
h_{\tr} &= - \frac{\left(4 + b_{\on}(6 + \chi_{\n}^2 r_{\h})\right) 
\left(2(-2 + b_{\on})\pi r_{\h}^2 + Q_0^2 \kappa\right)}{2\pi r_{\h}^5 \left(-2 + b_{\on}(-3 + \chi_{\n}^2 r_{\h})\right) \kappa} \, , &
\end{flalign}

\begin{flalign}
h_{4} &=- \frac{2(-2 + b_{\on})\pi r_{\h}^2 \left(4 + 3b_{\on}(2 + \chi_{\n}^2 r_{\h})\right) 
+ Q_0^2 \left(6 + b_{\on}(9 + 2\chi_{\n}^2 r_{\h})\right) \kappa}{8\pi r_{\h}^4 \left(-2 + b_{\on}(-3 + \chi_{\n}^2 r_{\h})\right) \kappa} \, ,&
\end{flalign}

\begin{flalign}\label{hu5v}
h_{5} &=\pm\frac{ \sqrt{ -2(-2 + b_{\on})\pi r_{\h}^2 - Q_0^2 \kappa }}{ \sqrt{\pi r_{\h}^3 \left( 2 + b_{\on}(3 + \psi_{\n}^2 r_{\h}) \right) }}\, , &
\end{flalign}

\begin{flalign}
h_{6} &=\frac{2(-2 + b_{\on}) \pi r_{\h}^2 \left(4 + b_{\on}(6 + \psi_{\n}^2 r_{\h})\right) 
+ (2 + 3b_{\on}) Q_0^2 \kappa}{8\pi r_{\h}^4 \left(2 + b_{\on}(3 + \psi_{\n}^2 r_{\h})\right) \kappa} \, ,&
\end{flalign}

\begin{flalign}\label{hu7v}
h_{7} &=- \frac{\left(-4 + b_{\on}(-6 + \psi_{\n}^2 r_{\h})\right) \left(2(-2 + b_{\on})\pi r_{\h}^2 + Q_0^2 \kappa\right)}{2\pi r_{\h}^5 \left(2 + b_{\on}(3 + \psi_{\n}^2 r_{\h})\right) \kappa} \, , &
\end{flalign}

\begin{flalign}
h_{8} &=\frac{-2(-2 + b_{\on}) \pi r_{\h}^2 \left(-4 + 3b_{\on}(-2 + \psi_{\n}^2 r_{\h})\right) 
+ Q_0^2 \left(6 + b_{\on}(9 - 2\psi_{\n}^2 r_{\h})\right) \kappa}{8\pi r_{\h}^4 \left(2 + b_{\on}(3 + \psi_{\n}^2 r_{\h})\right) \kappa} \, . &
\end{flalign}
\end{subequations}

\end{document}